\documentclass[12pt,a4paper,english]{article}
\usepackage{jheppub}
\usepackage{amsfonts}
\usepackage{mathtools}
\usepackage{braket}
\usepackage{multirow}
\usepackage[T1]{fontenc}
\usepackage[latin9]{inputenc}
\usepackage{color}
\usepackage{amsmath}
\usepackage{esint}
\usepackage{amssymb}
\usepackage{graphicx}

\usepackage{axodraw4j}
\usepackage{pstricks}
\usepackage{color}
\usepackage{bbold}

\newcommand{\be}{\begin{equation}}
\newcommand{\ee}{\end{equation}}
\newcommand{\bea}{\begin{eqnarray}}
\newcommand{\eea}{\end{eqnarray}}

\def\cD{{\cal D}}
\def\cF{{\cal F}}
\def\cO{{\cal O}}
\def\cP{{\cal P}}

\newcommand{\Tr}{\ensuremath{\operatorname{Tr}}}

\usepackage{babel}

\title{Quantum Gravity signatures \newline in the Unruh effect}

\author{Natalia Alkofer,}
\author{Giulio D'Odorico,}
\author{Frank Saueressig,}
\author{Fleur Versteegen}
\affiliation{
Institute for Mathematics, Astrophysics and Particle Physics (IMAPP),\\
Radboud University Nijmegen, Heyendaalseweg 135, 6525 AJ Nijmegen, The Netherlands
}
\emailAdd{n.alkofer@science.ru.nl}
\emailAdd{g.dodorico@science.ru.nl}
\emailAdd{f.saueressig@science.ru.nl}
\emailAdd{versteegen@science.ru.nl}

\abstract{We study quantum gravity signatures emerging from phenomenologically motivated multiscale models, spectral actions, and Causal Set Theory within the detector approach to the Unruh effect. We show that while the Unruh temperature is unaffected, Lorentz-invariant corrections to the two-point function leave a characteristic fingerprint in the induced emission rate of the accelerated detector. Generically, quantum gravity models exhibiting dynamical dimensional reduction exhibit a suppression of the Unruh rate at high energy while the rate is enhanced in Kaluza-Klein theories with compact extra dimensions. We quantify this behavior by introducing the ``Unruh dimension'' as the effective spacetime dimension seen by the Unruh effect and show that it is related, though not identical, to the spectral dimension used to characterize spacetime in quantum gravity. We comment on the physical origins of these effects and their relevance for black hole evaporation.
}


\begin{document}
\maketitle


\section{Introduction}
\label{intro}

Dimensional flows are a feature commonly encountered in virtually all approaches to quantum gravity and quantum gravity inspired models \cite{Carlip:2009kf,Carlip:2012md}. The most prominent example of a dimensional flow occurs in Kaluza-Klein theories where the dimensionality of spacetime increases below the compactification scale. An even more intriguing phenomenon of this form is dynamical dimensional reduction where a specific dimensionality of spacetime decreases at short distances. The prototypical example for this mechanism is provided by Causal Dynamical Triangulations \cite{Ambjorn:2012jv} where a random walk sees a two-dimensional spacetime at short distances while long  walks exhibit a four-dimensional behavior \cite{Ambjorn:2005db}. Similar features are encountered in Asymptotic Safety \cite{Lauscher:2005qz,Reuter:2011ah,Rechenberger:2012pm,Calcagni:2013vsa,Litim:2006dx}, Loop Quantum Gravity \cite{Modesto:2008jz,Caravelli:2009gk,Magliaro:2009if,Calcagni:2013dna,Calcagni:2014cza,Ronco:2016rtp}, Ho\v{r}ava-Lifshitz gravity \cite{Sotiriou:2011mu,Sotiriou:2011aa}, Causal Set Theory \cite{Eichhorn:2013ova,Carlip:2015mra,Belenchia:2015aia},  $\kappa$-Minkowski space \cite{Benedetti:2008gu,Anjana:2015ios,V.:2015msa},  non-commutative geometry \cite{Kurkov:2013kfa,Alkofer:2014raa}, non-local gravity theories \cite{Modesto:2011kw,Modesto:2015ozb}, minimal length models \cite{Padmanabhan:2015vma}, and based on the Hagedorn temperature seen by a gas of strings \cite{Atick:1988si}.

The indicator commonly used to study dimensional flows is the spectral dimension. The (typically Euclidean) quantum spacetime is equipped with an artificial diffusion process for a test particle. One then studies the return probability $P_\sigma$ of the particle as a function of the diffusion time $\sigma$. The mathematical definition of the spectral dimension is  obtained in the limit of infinitesimal diffusion time
\be\label{def:spectraldimension}
d_s = -2 \, \lim_{\sigma \rightarrow 0}  \frac{d \ln P_\sigma}{d \ln \sigma} \, . 
\ee
On a manifold the spectral dimension agrees with the topological dimension $d$.
In the context of quantum gravity where the properties of the underlying spacetime may depend on the length scales probed by the diffusing particle, it is useful to define a generalized spectral dimension $D_s(\sigma)$ where the limit $\sigma \rightarrow 0$ is omitted. The most common behavior of $D_s(\sigma)$ encountered in quantum gravity interpolates between $D_s = 4$ on macroscopic scales and $D_s = 2$ at short distances. This observation has also triggered the investigation of multi-scale geometries serving as a phenomenological model of quantum gravity inspired spacetimes \cite{Calcagni:2012rm}. 

The spectral dimension bears a close relation to the two-point correlation function $\widetilde G$ of the diffusing particle. For a massless scalar particle propagating on a four-dimensional Euclidean space one has $\widetilde{G} = p^{-2}$, which leads to a scale-independent spectral dimension $D_s = 4$. Non-trivial $D_s$-profiles are created if the two-point correlation function acquires an anomalous dimension. Based on this close connection, the interpretation of the spectral dimension as the Hausdorff dimension of the momentum space has been advocated in \cite{Amelino-Camelia:2013gna}. Note that a non-trivial spectral dimension does not necessarily involve the breaking of Lorentz invariance, since $\widetilde{G}(p^2)$ may be a function of the momentum four-vector squared and thus a Lorentz invariant quantity.
However, this function can in principle have more general forms than those allowed in a local quantum field theory.
One relevant example is a two-point function arising in a nonlocal field theory, defined as a theory
whose equations of motion have an infinite number of derivatives.
This form is ubiquitous in Causal Set studies \cite{Aslanbeigi:2014zva}.

The fictitious nature of the diffusion process underlying the spectral dimension then raises the crucial question whether the flow of the spectral dimension can be seen in a physical observable quantity. The main goal of this paper is to explicitly demonstrate that this is indeed the case: the non-trivial momentum profiles leave an imprint in the Unruh effect  felt by an accelerated detector. More precisely, the effective dimension of spacetime seen by the Unruh detector is determined by the spectral dimension.

The Unruh effect \cite{Fulling:1972md,Davies:1974th,Unruh:1976db} (also see \cite{Crispino:2007eb,Birrell:1982ix,Takagi:1986kn} for reviews) is one of the most intriguing phenomena occurring within quantum field theory in Minkowski space. Essentially, it predicts that to an accelerated observer (Rindler observer) the Minkowski vacuum appears as a thermal state whose temperature is proportional to the acceleration parameter.
This acceleration radiation can leave imprints in a variety of phenomenological contexts:
for instance in the transverse polarization of electrons and positrons in particle storage rings (Sokulov-Ternov effect)
\cite{Akhmedov:2006nd,Akhmedov:2007xu},
at the onset of quark gluon plasma formation due to heavy ions collisions \cite{Kharzeev:2005iz},
on the dynamics of electrons in Penning traps, of ultra-intense lasers, and atoms in microwave cavities
(see \cite{Crispino:2007eb} and references therein), or in the Berry phase acquired by the accelerated detector \cite{MartinMartinez:2010sg}.
Recently it has also been shown that the low energy signatures of Unruh radiation are very sensitive
to high energy nonlocality \cite{Belenchia:2016sym}.

On theoretical grounds the Unruh effect can be derived by defining creation and annihilation operators with respect to the positive and negative frequency modes associated with the Minkowski and Rindler space and relating them through a Bogoliubov transform, see e.g. \cite{Mukhanov:2007zz} for a pedagogical exposition. The origin of the thermal spectrum is essentially geometrical, in the sense that it depends solely on the presence of an horizon in the Rindler frame. As a geometric effect, the Unruh temperature is insensitive to the specific form of the Lagrangian or the interactions under consideration and thermality of the spectrum is essentially ensured by Lorentz invariance \cite{Unruh:1983ac}. We show that this also holds for the broad class of quantum gravity corrections considered in this work.\footnote{For similar studies in
the context of anisotropic dispersion relations and a minimal length scale see \cite{Rinaldi:2008qt,Nicolini:2009dr,Agullo:2010iq,Husain:2015tna} }
While not affecting the thermal nature of the Unruh radiation, quantum gravity induced modifications 
of the two-point function affect the profile functions multiplying the thermal distribution in a more or less radical way.

In order to make the connection between dimensional flows and modifications in the Unruh effect as close as possible, we follow the detector approach \cite{Agullo:2010iq}. The central idea is to consider a detector made from a two-level system with an upper, excited state $2$ and a lower state $1$ being separated by the energy $\Delta E \equiv E_2 - E_1 > 0$ coupled to a scalar field. The transition probabilities induced by the scalar can be expressed in terms of the positive-frequency Wightman function of the Minkowski vacuum state.  The emission rates of the detector
can be computed by evaluating a Fourier transform  of the two-point function along the worldline of an accelerated observer. For a standard massless scalar field, it is then rather straightforward to show that 
the Green's function evaluated on the worldline satisfies a Kubo-Martin-Schwinger (KMS) condition where the periodicity 
in Euclidean time depends on the properties of the worldline only. The resulting Unruh temperature is
proportional to the acceleration $a$. This setup also makes clear that 
corrections to the two-point functions, 
e.g. induced by quantum fluctuations at small scales, may leave their fingerprints in the transition rate of the Unruh detector. Both, a dynamical dimensional flow and corrections to the transition rate, can be traced back to the same source: a non-trivial momentum dependence of the two-point function.

In this work we will focus on the asymptotic structure of the detector-induced emission rates in a fixed Minkowski background.\footnote{Throughout the work
we will not take into account effects related to the ``switching function'' $\chi$, which controls the time dependence of the detector 
coupling strength, see \cite{MartinMartinez:2012th,Alhambra:2013uja} for details.}
We will show that different types of dimensional flows leave distinct  signatures in the detector rates.
In particular, in the case of dimensional reduction at high energies, one finds a suppression of the rates,
whereas for a dimensional \emph{enhancement} at high energies, as in Kaluza-Klein models,
the rate increases.
Since the transition probability of the Unruh detector is clearly a signature which is observable at least in principle, we expect that it can be used to make phenomenological predictions from quantum gravity allowing a direct comparison between various approaches.

The rest of the work is organized as follows. Sect.\ \ref{sect.2} briefly reviews the detector approach to the Unruh effect. Dimensional flows entail specific modifications of the two-point correlation functions entering into the detector approach and we derive the master formula capturing the resulting corrections to the Unruh effect in Sect.\ \ref{sect.3}. 
In Sect.\ \ref{sect.3b} we define the Unruh dimension as the effective dimension  seen by the detector and relate it to the spectral dimension.
In Sect.\ \ref{sect.4} we apply this formula to specific examples taken from phenomenologically motivated multiscale models (Sect.\ \ref{sect.41}), Kaluza-Klein theory (Sect.\ \ref{sect.43}), spectral actions (Sect. \ref{sect.44}), and Causal Set Theory (Sect.\ \ref{sect.45}). We close with a brief discussion of our findings in Sect.\ \ref{sect.5}. For completeness, technical details are relegated to two appendices.

\section{Rates from correlators}
\label{sect.2}
In this work we follow the detector approach to the Unruh effect  \cite{Birrell:1982ix,Unruh:1983ms,HawkingIsrael:1979,Agullo:2010iq}. The advantage of this approach is that it considers observable quantities, namely emission and absorption rates of the accelerated detector. The response of the accelerated detector then indicates that it is immersed in a thermal bath of particles.  
This framework is ideally suited for studying
 corrections to the Unruh effect by using effective two-point correlation functions incorporating quantum gravity effects. We first review
 the formalism following \cite{Agullo:2010iq} before applying it to  dimensional flows in Sects.\ \ref{sect.3}, \ref{sect.3b}, and \ref{sect.4}.

\subsection{Particle detectors and two-point functions}
The simplest model of a particle detector \cite{Unruh:1983ms,HawkingIsrael:1979,Agullo:2010iq} is a quantum mechanical
system with two internal energy states $|E_2\rangle$ and $|E_1\rangle$, 
with energies $E_2>E_1$. 
The detector moves along a worldline $x(\tau)$ parameterized by the detector's proper time $\tau$ and interacts with a scalar field $\Phi(x)$ by absorbing or emitting its quanta. 
The coupling of $\Phi$ to the detector   is  modeled by a monopole moment operator $m(\tau)$ 
acting on the internal detector eigenstates through the Lagrangian
\be 
\label{detectorPhi} 
{L}_I=  g \ m(\tau)\Phi(x(\tau)) \, .
\ee

We will consider in the following the two cases of a detector moving inertially in Minkowski space,
and one moving along a uniformly accelerated trajectory, which defines the Rindler space (see Appendix A).
Let us denote the Minkowski vacuum by $|0_M\rangle$, the Rindler vacuum by $|0_R\rangle$, and  the one-particle state of the field $\Phi$ with spatial momentum $\vec k$ by $|\vec{k}\rangle $. 
There are three possible processes giving a non-zero rate.
Following the nomenclature used in \cite{Agullo:2010iq}
we can also give them a thermodynamic interpretation, since it will turn out that Rindler correlators are thermal.
First, the inertial detector can be in the excited state with energy $E_2$.
This is a spontaneous emission process and corresponds to the transition
$|E_2\rangle |0_M\rangle \  \to  |E_1\rangle |\vec{k}\rangle $ for an observer comoving with the detector. 
Second, the accelerating detector can be in the excited state with energy $E_2$.
This is an induced emission process and instead corresponds to the transition 
$|E_2\rangle |0_R\rangle \  \to  |E_1\rangle |\vec{k}\rangle $
for an inertial observer in Minkowski space (or equivalently $|E_2\rangle |0_M\rangle \  \to  |E_1\rangle |\vec{k}\rangle $ for
an \emph{accelerating} one). 
Finally, an accelerating detector in the ground state $E=E_1$ corresponds to
absorption, or the transition $ |E_1\rangle | 0_M \rangle \to |E_2\rangle |\vec{k} \rangle $.
Notice that the term absorption here is meant purely as an analogy with two state systems,
since the one-particle state $|\vec{k}\rangle$ still appears as a final state.


The transition probability can be expressed in terms of the two-point function of the field.
To first order in time-dependent perturbation theory, the amplitude
for the detector-field interaction takes the form
\be
{\cal A}(\vec{k}) = ig \langle E_f|m(0)| E_i\rangle \int d\tau e^{i(E_f-E_i) \tau} \langle \vec{k}|\Phi(x(\tau))|0_M\rangle \ . 
\ee
The transition probability is the square of the amplitude, integrated over all possible final states
\be
P_{i \to f} = \int d^3k | {\cal A}(\vec{k})| ^2 \, .
\ee
For $E_f = E_1$ and $E_i = E_2$ this gives 
 the total, spontaneous plus induced, emission probability.

The field $\Phi$ can be expanded in its normal mode basis, according to the choice of vacuum.
If we define the annihilation operators in Minkowski space as $a_{\vec{k}} |0_M\rangle =0$,
and those in Rindler space (we work implicitly in the right wedge) as $b_{\vec{k}} |0_R\rangle =0$,
then the field has the expansions:
\be
\Phi(x) = \int d^3k \left(u_{\vec{k}} a_{\vec{k}} + u^{*}_{\vec{k}} a^{\dagger}_{\vec{k}}\right) 
= \int d^3k \left(v_{\omega\vec{k}_{\bot}} b_{\omega\vec{k}_{\bot}} + v^{*}_{\omega\vec{k}_{\bot}} b^{\dagger}_{\omega\vec{k}_{\bot}}\right) \ .
\ee
We used the notation $\vec{k}_{\bot}=(k_y,k_z)$, these coordinates are left untouched by the Rindler coordinate transformation.
Here the mode functions in the Minkowski basis are
\be
u_{\vec{k}}= \frac{1}{\sqrt{2(2\pi)^3w}} e^{-i(wt-\vec{k}\vec{x})} \, ,
\ee
where $w \equiv \sqrt{\vec k^2 + m^2}$,
whereas in the Rindler basis with coordinates $(\tau, \xi, \vec{x}_{\bot})$ they are given in terms of a modified Bessel function $K_\nu(x)$ as
\cite{Crispino:2007eb}
\be
\label{vRind}
v_{\omega\vec{k}_{\bot}} = \left[ \frac{\sinh(\pi\omega/a)}{4\pi^2 a} \right]^{1/2} 
K_{i\omega/a}\left( \frac{\sqrt{\vec{k}_{\bot}^2 + m^2}}{a} e^{a\xi} \right)
e^{-i(\omega \tau-\vec{k}_{\bot}\cdot\vec{x}_{\bot})} \, .
\ee
The sum over all possible one-particle states needed to obtain the transition probabilities leads to a sum over modes
$\sum_{\vec{k}}u_{\vec{k}}(x_1)u^{*}_{\vec{k}}(x_2)$.
Upon using the completeness of states this gives rise to the two-point function for the Minkowski vacuum.
Defining $C \equiv g^2 |\langle E_f |m(0)|E_i\rangle|^2$, one finds
\bea
P_{i \to f} &=& C  \int d^3k
\int_{-\infty}^{\infty}d\tau_1 \int_{-\infty}^{\infty}d\tau_2 \, e^{i (E_f - E_i) (\tau_1-\tau_2)}
\langle \vec{k}|\Phi(x(\tau_1))|0_M\rangle \langle 0_M|\Phi(x(\tau_2))|\vec{k}\rangle \nonumber
 \\ 
&=& C
\int_{-\infty}^{\infty}d\tau_1 \int_{-\infty}^{\infty}d\tau_2 \, e^{i (E_f - E_i) (\tau_1-\tau_2)}
\langle 0_M|\Phi(x(\tau_2)) \Phi(x(\tau_1))|0_M\rangle .
\eea

Performing the integration over all the final states first, the expression for the transition probabilities then becomes \cite{Agullo:2010iq}
\be\label{Pif} 
P_{i \to f} = C \, F(\Delta E) \ , 
\ee
where
$F(\Delta E)$ is the so-called response function
\be 
\label{FE1}
F(\Delta E)=
\int_{-\infty}^{\infty}d\tau_1 \int_{-\infty}^{\infty}d\tau_2 \, 
e^{-i ( E_f - E_i) \Delta \tau} G_M(\Delta \tau -i\epsilon) \ .  
\ee 
Here $\Delta \tau \equiv \tau_1 -\tau_2$ (from now on the limit $\epsilon\to 0^+$ is understood). The response function is essentially given by the Fourier transform of the Wightman two-point function $G_M(\Delta \tau -i\epsilon)$ evaluated on the detector's trajectory.

In the following we will be interested in the emission case, with $E_i=E_2$ and $E_f=E_1$ and $\Delta E \equiv E_2 - E_1$ is taken positive by definition.
For the case of the detector undergoing constant acceleration the total transition probability \eqref{Pif} contains contributions from spontaneous and induced emission.
Subtracting the spontaneous emission probability
one arrives at the following formula for the induced emission response function
\be
\label{parkereq}
F_I(\Delta E)=
\int_{-\infty}^{\infty}d\tau_1d\tau_2 \, 
e^{i \Delta E \Delta \tau} \left[ G_M(\Delta \tau -i\epsilon) - G_{R}(\Delta \tau -i\epsilon) \right] \, . 
\ee
Here $G_{M}$ is the vacuum (Wightman) two-point function for an observer
on the accelerated trajectory in the Minkowski vacuum, 
\be
G_{M}\left(x,x^{\prime}\right)=\left\langle 0_{M}\right|\Phi\left(x\right)\Phi\left(x^{\prime}\right)\left|0_{M}\right\rangle \, ,  
\ee
and $G_{R}$ is the vacuum two-point function of an accelerated
observer in the Rindler vacuum,
\be
G_{R}\left(x,x^{\prime}\right)=\left\langle 0_{R}\right|\Phi\left(x\right)\Phi\left(x^{\prime}\right)\left|0_{R}\right\rangle \, . 
\ee

Practically, it is convenient to work with the induced transition rate per unit time given by
\be 
\dot P_{i \to f} = g^2 \, |\langle E_f |m(0)|E_i\rangle|^2 \, \dot F_I(\Delta E) \ , 
\ee
with 
\be
\label{parkereq2}
\dot F_I(\Delta E) =
\int_{-\infty}^{+\infty}d\Delta \tau \, 
e^{i \Delta E \Delta \tau} \, \left[ G_M(\Delta \tau -i\epsilon) - G_{R}(\Delta \tau -i\epsilon) \right] \, . 
\ee
This equation is the relation between physical rates and two-point functions that we will use in the following. In order to ease our notation we will set $\Delta \tau = \tau$ and $\Delta E = E$ from now on.

The Wightman function for a massive scalar field with mass $m$ in Minkowski space entering into eq.\ \eqref{parkereq2} is given by
\be\label{Wightmanfct}
G_+(\vec{x},t) = -i \int \frac{d^3 \vec p}{(2\pi)^3} \oint_{\gamma_+} \frac{dp^0}{2\pi} \, \widetilde{G}(p^2) \, e^{i\vec{p}\cdot\vec{x} - ip^0 t} \, ,
\ee
where
\be\label{genfct}
\widetilde{G}(p^2) = \frac{1}{p^2 - m^2} = \frac{1}{(p^0 + \sqrt{\vec p^2 + m^2})(p^0 - \sqrt{\vec p^2 + m^2})} \, . 
\ee
The contour $\gamma_+$ encircles the first order pole located at $p^0 = \sqrt{\vec p^2 + m^2}$. Carrying out the Fourier integral the 
 positive-frequency Wightman function in Minkowski space is given by (see, e.g., \cite{Birrell:1982ix})
\be\label{massiveWightman}
G_M (x, x^\prime) = - \frac{i m}{4\pi^2} 
\frac{K_1 \left( im \sqrt{\left( t - t^{\prime} -i\epsilon \right)^2 - \left( \vec{x} - \vec{x}^{\prime} \right)^2 } \right)}{\sqrt{\left( t - t^{\prime} -i\epsilon \right)^2 - \left( \vec{x} - \vec{x}^{\prime} \right)^2 }} \, . 
\ee
Here $K_1$ is the modified Bessel function of the second kind. 
In the massless limit \eqref{massiveWightman} reduces to the Wightman function of a massless scalar field in position space \cite{Birrell:1982ix,Agullo:2010iq}
\be\label{masslessWightman}
G_M (x, x^\prime) = - \frac{1}{4\pi^2} \, 
\frac{1}{\left( t - t^{\prime} -i\epsilon \right)^2 - \left( \vec{x} - \vec{x}^{\prime} \right)^2 } \, . 
\ee

The Wightman function in Rindler space is just the same evaluated on the
worldline of the uniformly accelerated detector
\be\label{rindlertrajectory}
t = a^{-1} \, \sinh(a \tau) \, , \quad
 x = a^{-1} \, \cosh(a \tau) \, , \quad
 y = 0 \, , \quad z = 0 \, . 
\ee
For a thermal system, the induced emission probability coincides with the absorption probability.
We can then turn to the proof that the Minkowski vacuum corresponds 
to a thermal state when probed by an accelerated detector.

\subsection{Emergence of thermality}
The advantage of working with Rindler geometry is that it shows how the Unruh thermal spectrum
is a geometric effect. It arises for a generic Lorentz-invariant matter theory simply because 
of the properties of the Rindler frame (see App.\ \ref{App.A} for more details).
 
Consider a generic Lorentz invariant Green's function $G_M(x,x^{\prime})=G_M(x-x^{\prime})$ for an interacting theory in Minkowski space.
When evaluated on the worldline \eqref{rindlertrajectory} of a uniformly accelerated observer, 
it will be a function of the Rindler coordinates
$(\vec{x}_\bot, \tau)$ and $(\vec{x}^{\prime}_\bot, \tau^{\prime})$.
Since the theory is Lorentz invariant, $G_M$ can only depend on $(x-x^{\prime})^2$.
Using the relation 
\be
\begin{split}
 (t-t^{\prime})^2 - (x-x^{\prime})^2  = & \, a^{-2} \left[ ( \sinh a\tau - \sinh a\tau ^{\prime})^2 - 
 (\cosh a\tau - \cosh a\tau ^{\prime})^2 \right]  \\
 = & \, 2 a^{-2} \left(  \cosh (a\Delta\tau) -1 \right)  \,,
\end{split}
\ee
with $\Delta\tau=\tau - \tau^{\prime}$, the Rindler Green's function has a $\tau$ dependence of the form $G_R(\cosh a\Delta\tau)$.
Focusing for simplicity on $\tau^{\prime}=0$, a Wick rotation $t=it_E$ will induce, through $t=a^{-1} \sinh a\tau$, a corresponding Wick rotation
in Rindler time, $\tau = i \tau_E$. But this then means that a general Rindler two-point
function will be periodic in Rindler time, since $G_R(\cosh a\tau) \to G_R^{(E)}(\cos a\tau_E)=G_R^{(E)}(\cos (a\tau_E+2\pi))$.
We thus see\footnote{
There is a subtlety in the Wick rotation when working with Wightman functions.
Due to the different domains of analyticity of $G_+$ and $G_-$ in the complex $\tau$-plane,
one actually identifies $G_E(\tau_E)=G_+(i \tau_E)$ for $-2\pi<\tau_E<0$
and $G_E(\tau_E)=G_-(i \tau_E)$ for $0<\tau_E<2\pi$.
This is responsible for the change of sign of $\tau$ in the KMS condition.}
that the periodicity $\beta=2\pi/a$ implies a temperature $T=a/2\pi$.

Undoing the Wick rotation we obtain the KMS condition in the form (with obvious change of notation)
\be
G_R(\tau) = G_R(-\tau-i\beta) \, . 
\ee
This can be put in another equivalent form, which is more natural when dealing with detector rates \cite{Takagi:1986kn}.
Since the rate is a Fourier transform of the Wightman function, assuming that
$G_R(\tau)$ is analytic in the strip $-\beta<{\rm Im}\tau<0$, we have
\bea
\dot{F}(E) &=& \int_{-\infty}^{+\infty} d\tau e^{-iE\tau} G_R(\tau-i\epsilon) \nonumber \\
&=& \int_{-\infty}^{+\infty} d\tau e^{-iE(\tau-i\beta+2i\epsilon)} G_R(\tau-i\beta+i\epsilon) \nonumber \\
&=& e^{-(\beta-2\epsilon)E} \int_{-\infty}^{+\infty} d\tau e^{iE\tau} G_R(\tau-i\epsilon) \, . 
\eea
Here in the second line we made use of the analyticity assumption to push down the contour
in the complex $\tau$-plane by $i(\beta-2\epsilon)$, and in the third line we changed variable of integration to $-\tau$.
Taking $\epsilon$ to zero, the KMS condition becomes
\be
\label{KMS}
\dot{F}(E) = e^{-\beta E} \dot{F}(-E) \, . 
\ee
This relation can also be derived 
 directly in the free massive case from
the parity properties of the integrands appearing in the rates.\footnote{We thank J. Louko for pointing this out to us.}
A general proof of the KMS condition for an interacting field theory in any dimension 
was given in \cite{Sewell}.

The Unruh temperature is thus only determined by the Euclidean periodicity,
and is protected against corrections as long as the Lorentz invariance of $G_M$ is preserved.
In particular, if one computes the average number density $\langle n \rangle$ in Rindler space
from thermal considerations alone, one can obtain the usual Planckian distribution
with temperature $T=a/2\pi$.
As a simple illustration of this fact, in the next section we will derive
the Planckian thermal spectrum for a massive scalar field,
showing as a byproduct that the temperature is independent of the mass.

\subsection{Detector response for massive scalars}
\label{sect.23}


Let us start with the massless case for illustration purposes.
In this case the rate integral can be computed directly, by closing the contour 
with a large semicircle in the upper complex-$\tau$ half-plane.
The contour can be deformed to infinity into a sum over the infinite number
of poles of the integrand located along the imaginary axis.
This sum then gives rise to the Matsubara thermal sum that generates
the Planckian thermal factor.

One can also recover the same result from the KMS condition.
With reference to the nomenclature previously introduced,
let us call $\dot{F}_A$ the absorption rate and $\dot{F}_E$ the emission rate.
This last one is the sum of spontaneous and induced emission, $\dot{F}_E = \dot{F}_S + \dot{F}_I$.
From the derivation of the formulas for the detector rates in Sect.\ 2.1,
one immediately finds that $\dot{F}_A(-E) = \dot{F}_E(E)$.
This is ensured by the fact that the one-particle state $|\vec{k}\rangle $
always appears as a final state, and thus the Wightman function has the same frequency for both processes.
The difference then just amounts to the sign of the Fourier exponential term.
Using the KMS condition eq.\ (\ref{KMS}), this gives
\be
\dot{F}_A(E) = e^{-\beta E}\dot{F}_A(-E) = e^{-\beta E}\dot{F}_E(E) = e^{-\beta E} [ \dot{F}_I(E) + \dot{F}_S(E) ] \,.
\ee
%
If the induced emission and absorption rates coincide
\be
\label{equal}
\dot{F}_A(E) = \dot{F}_I(E)
\ee
it follows that
\be
\label{induced}
\dot{F}_I(E) = \frac{\dot{F}_S(E)}{e^{\beta E}-1} \, . 
\ee
Thus one only needs to compute the spontaneous rate to obtain that for induced emission.
In the massless case this is easily computed to give $\dot{F}_S(E)=E/2\pi$.

Condition (\ref{equal}) unfortunately does not strictly hold for a free massive scalar field.
An explicit calculation in this case \cite{Takagi:1986kn} gives for the total rate
\bea
\label{4.1.11}
\dot{F} (E) &=& \int_{-\infty}^{+\infty}d\tau e^{-iE\tau} G_R(\tau-i\epsilon)
\nonumber \\
&=& 2\pi \int d^2 k_{\bot} \left| v_{\omega\vec{k}_{\bot}} \right|^2 \left[ \theta(E) N(E/a) + \theta(-E) (1+ N(|E|/a) ) \right] \,,
\eea
where
\be
N(x)= \frac{1}{e^{2\pi x}-1} \, . 
\ee

If we interpret the different terms in eq.\ (\ref{4.1.11}) following the language of Sect.\ 2.1,
the first term corresponds to the absorption case, while the second is the sum of induced emission
plus
the contribution from 
an accelerated detector in the Rindler vacuum, with $\xi=0$ and $\vec{x}_{\bot}=0$.
An explicit calculation of this term following \cite{Agullo:2010iq} (see eq. (3.11) in that reference) gives indeed
\be
\label{3.11m}
\dot{F}_S(E) = 2\pi \int d^2 k_{\bot} d\omega \left| K_{i\omega/a}\left( \frac{\sqrt{\vec{k}_{\bot}^2 + m^2}}{a} \right) \right|^2 \frac{\sinh(\pi\omega/a)}{4\pi^4 a}\delta(\omega-E)
\ee
which, using eq.\ (\ref{vRind}), precisely reproduces the "spontaneous" term in eq.\ (\ref{4.1.11}).
Unfortunately, this does not in general coincide with the true spontaneous rate,
defined as the rate of a detector at rest in Minkowski space.
Intuitively, the two notions should coincide, but in this case the difference lies
in the absence of a mass gap in Rindler space.

To show this, consider a detector at rest in the Minkowski vacuum, in general dimension $d$ \cite{Birrell:1982ix}.
The simplest way to compute the rate is to start from eq.\ \eqref{FE1} and substitute the explicit form of the two-point function:
\be
\label{response}
F(\Delta E)=
\int_{-\infty}^{\infty} d\tau_1 \int_{-\infty}^{\infty} d\tau_2 \, 
e^{ i \Delta E \Delta \tau} 
\int \frac{d^d k}{(2\pi)^d} \frac{1}{2\omega} 
e^{-	i \omega \left( t(\tau_1) - t^{\prime}(\tau_2) \right) +i \vec{k}\cdot \left( \vec{x}(\tau_1) - \vec{x}^{\prime} (\tau_2) \right)
} \, .
\ee
Inverting the $\tau$ and $k$ integrations we find
\bea
\label{spont}
\dot{F}_S(E) &=&
\int \frac{d^{d-1} k}{(2\pi)^{d-1}} \frac{1}{2 \sqrt{k^2+m^2}} 
\int_{-\infty}^{+\infty} d\tau e^{- i (\sqrt{k^2+m^2} - E) \tau}
\nonumber \\
&=& \frac{\pi^{\frac{d-1}{2}}}{\Gamma(\frac{d-1}{2})(2\pi)^{d-2}} \left(E^2-m^2 \right)^{\frac{d-3}{2}}\theta(E-m) \,.
\eea
%

\begin{figure}
	\begin{centering}
		\includegraphics[scale=.85]{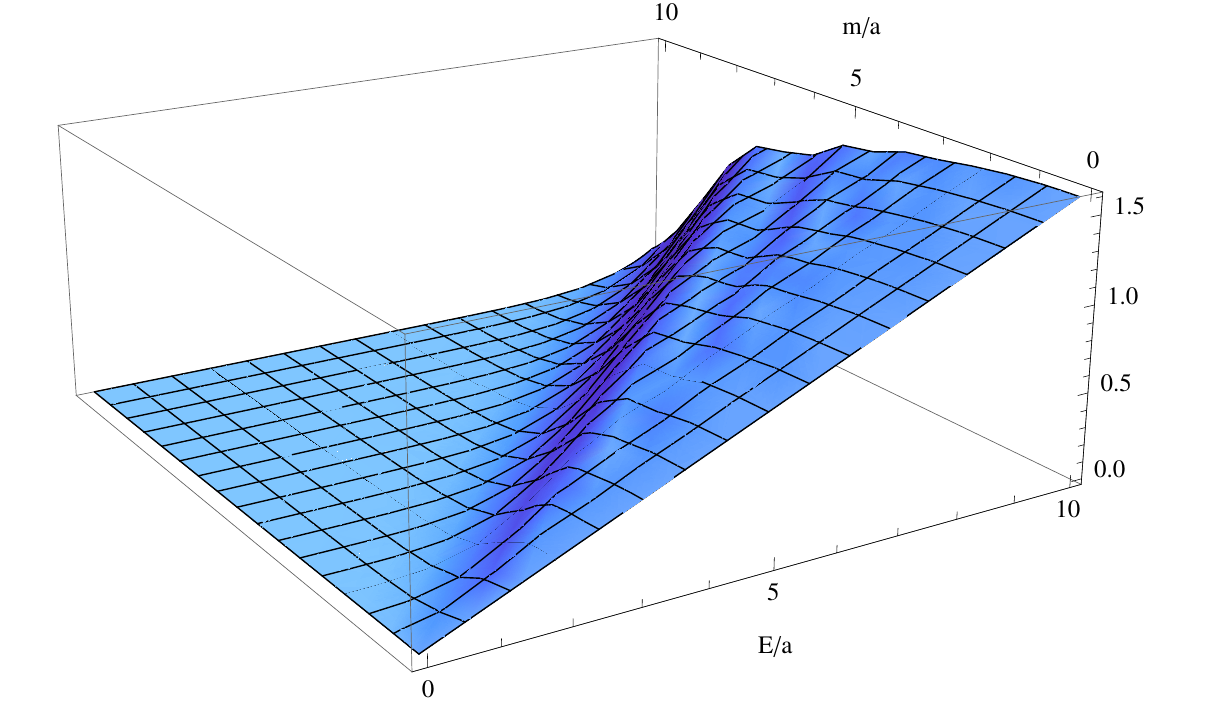}
		\caption{\label{gap} Numerical integration of eq. \eqref{3.11m}, as a function of the
		dimensionless ratios $E/a, m/a$.}
	\end{centering}
\end{figure}

The relations \eqref{3.11m} and \eqref{spont} coincide in the limit where $E\gg m$.
A crucial difference between the two results is that \eqref{spont} exhibits
a mass gap which is absent in \eqref{3.11m}.
The numerical integration of \eqref{3.11m}, displayed in Fig. \ref{gap},
shows that this expression well approximates \eqref{spont} when $E<m$.
Thus we will use this approximation in the sequel.
Incidentally, this also shows that condition \eqref{equal}, though not exact,
is approximately satisfied in the massive case.

Exploiting now relation (\ref{induced}), 
the induced rate function per unit time of the accelerated detector in $d=4$ becomes
\be\label{Fdotfinal}
\dot F = \frac{1}{2 \pi} \, \sqrt{E^2 - m^2} \, \theta(E-m) \, \frac{1}{e^{\frac{2\pi E}{a}} -1} \, .
\ee
The rate function constitutes the main result of this subsection. Taking the limit $m \rightarrow 0$, it agrees with the derivation for the massless case given in \cite{Birrell:1982ix,Agullo:2010iq}. The structure of \eqref{Fdotfinal} then motivates the definition of a profile function $\mathcal{F}(E)$
via
\be\label{defresf}
\dot F = \frac{1}{2 \pi} \, \mathcal{F}(E) \, \frac{1}{e^{\frac{2\pi E}{a}} -1} \, .
\ee
For a massless and massive scalar field obeying the Klein-Gordon equation one then has
\be
\mathcal{F}^{\rm massless}(E) = E \, , \qquad 
\mathcal{F}^{\rm massive}(E) = \sqrt{E^2 - m^2} \, \theta(E-m) \, . 
\ee
For general dimension the profile function is
\be
\label{rategen}
{\cal F}(E) = \frac{\pi^{\frac{d-1}{2}}}{\Gamma(\frac{d-1}{2})(2\pi)^{d-3}} 
\left(E^2-m^2 \right)^{\frac{d-3}{2}}\theta(E-m) \, . 
\ee
As we will show in the subsequent section, it is this profile function 
that actually carries information about quantum gravity corrections to
the Unruh rate.

As stressed before, the Planckian thermal factor is independent of the details of the field considered.
The fact that the mass dependence enters through the prefactor tells us that
the signatures of the fields involved will only be present in physical rates,
and not in number densities $\left\langle n \right\rangle$. 


\section{Master formulas for modified detector rates}
\label{sect.3}
In the presence of a dimensional flow, $\tilde G(p^2)$ entering into \eqref{Wightmanfct}, acquires a non-trivial momentum dependence.\footnote{As noted before, this does not necessarily entail the breaking of Lorentz symmetry since $\tilde G(p^2)$ may still be a Lorentz invariant function depending on the square of the momentum four-vector only.} 
It is useful to distinguish the two cases where $[\tilde G(p^2)]^{-1}$ is a polynomial in $p^2$ or given by a more general function with a finite number (typically one) of zeros in the complex $p^0$-plane. These two cases will be discussed in Sects.\ \ref{sect.31} and \ref{sect.32}, respectively.

\subsection{Detector rates from the Ostrogradski decomposition}
\label{sect.31}
We start by considering the case in which $\left[\widetilde{G}(p^2)\right]^{-1} \equiv {\cal P}_n (p^2)$ is an inhomogeneous polynomial of order $n$.
This covers the class of theories with a general quadratic effective Lagrangian ${\cal L} = \tfrac{1}{2} \phi \, {\cal P}_n(-\partial^2) \, \phi$ where ${\cal P}_n$ is a local function of the flat space d'Alembertian operator that admits a Taylor expansion around zero momentum. This comprises all local theories in which higher order corrections come in definite powers of momenta.
The limiting case $n\to\infty$ can also be considered. In this case the profile function ${\cal F}(E)$, eq.\ \eqref{defresf}, can be constructed from the Ostrogradski decomposition for a higher-derivative field theory.

The polynomial ${\cal P}_n (z)$ has $n$ roots, $\mu_i, i = 1,\ldots,n$ in the complex $z$-plane. It can then be factorized according to
\begin{equation}\label{poly1}
{\cal P}_n (z) = c \, \prod_{i=1}^{n} \left(z - \mu_i \right)
\end{equation}
where $c$ is a normalization constant. In order to connect to the case of a massive scalar field, the momentum space propagator is decomposed according to
\begin{equation}\label{poly2}
\left[ {\cal P}_n (z) \right]^{-1} = \frac{1}{c} \, \sum_{i=1}^n \frac{A_i}{\left(z - \mu_i \right)}
\end{equation}
where the coefficients $A_i$ are functions of the roots $\mu_i$. Assuming that $z \not = \mu_i$, eqs.\ \eqref{poly1} and \eqref{poly2} can be multiplied to obtain the condition
\begin{equation}\label{aconst}
\sum_{i=1}^n A_i \, \prod_{j\neq i}\left(z - \mu_j \right) =1 \, . 
\end{equation}
This condition must hold for any value $z \not = \mu_i$. Since the left-hand-side is a polynomial in $z$ of order $n-1$, \eqref{aconst} gives
rise to $n$ equations determining the coefficients $A_i$. Defining the vector
${\cal Z} \equiv \left[1,z,\ldots,z^{n-1}\right]$ and introducing the coefficient matrix $\cal C$ via ${\cal C}_{ij} \, {\cal Z}_j \equiv \prod_{j\neq i}\left(z - \mu_j \right)$, eq.\ \eqref{aconst} entails
\begin{equation}
\sum_{i=1}^{n}  A_i  \, {\cal C}_{ij} = \, \delta_{1j} \, , 
\end{equation}
where $\delta_{ij}$ is the Kronecker symbol. This equation can be solved for $A_i$ if  ${\cal C}$ is invertible, i.e. $\det{\cal C}\neq 0$.
The general condition for the two-point function to be factorizable then
is $\mu_i \not = \mu_j, i \not = j$, i.e., all roots of the polynomial have order one.

 Assuming that these conditions are met, the solution for the $A_i$ is given by the first row of the inverse matrix ${\cal C}$, $A_i = \left( {\cal C}^{-1}\right)_{1i}$.
The explicit solution for the $A_i$ is then given by
\be\label{Aisol}
A_i = \left( {\prod_{j \not = i} (\mu_i - \mu_j)} \right)^{-1} \, . 
\ee
For future reference, it is convenient to give the coefficients $A_i$ entering   the decomposition \eqref{poly2} for the cases $n=2$ and $n=3$ explicitly. For $n=2$,
\be
A_1 = \frac{1}{\mu_1 - \mu_2} \, , \quad 
A_2 = \frac{1}{\mu_2 - \mu_1}
\, ,  
\ee
while for $n=3$ one has 
\be
A_1 = \frac{1}{(\mu_1 - \mu_2)(\mu_1 - \mu_3)} \, , \;
A_2 = \frac{1}{( \mu_2- \mu_1)(\mu_2 - \mu_3)} \, , \;
A_3 = \frac{1}{(\mu_3 - \mu_1)(\mu_3 - \mu_2)} \, .
\ee

\begin{figure}
	\begin{centering}
		\includegraphics[scale=.35]{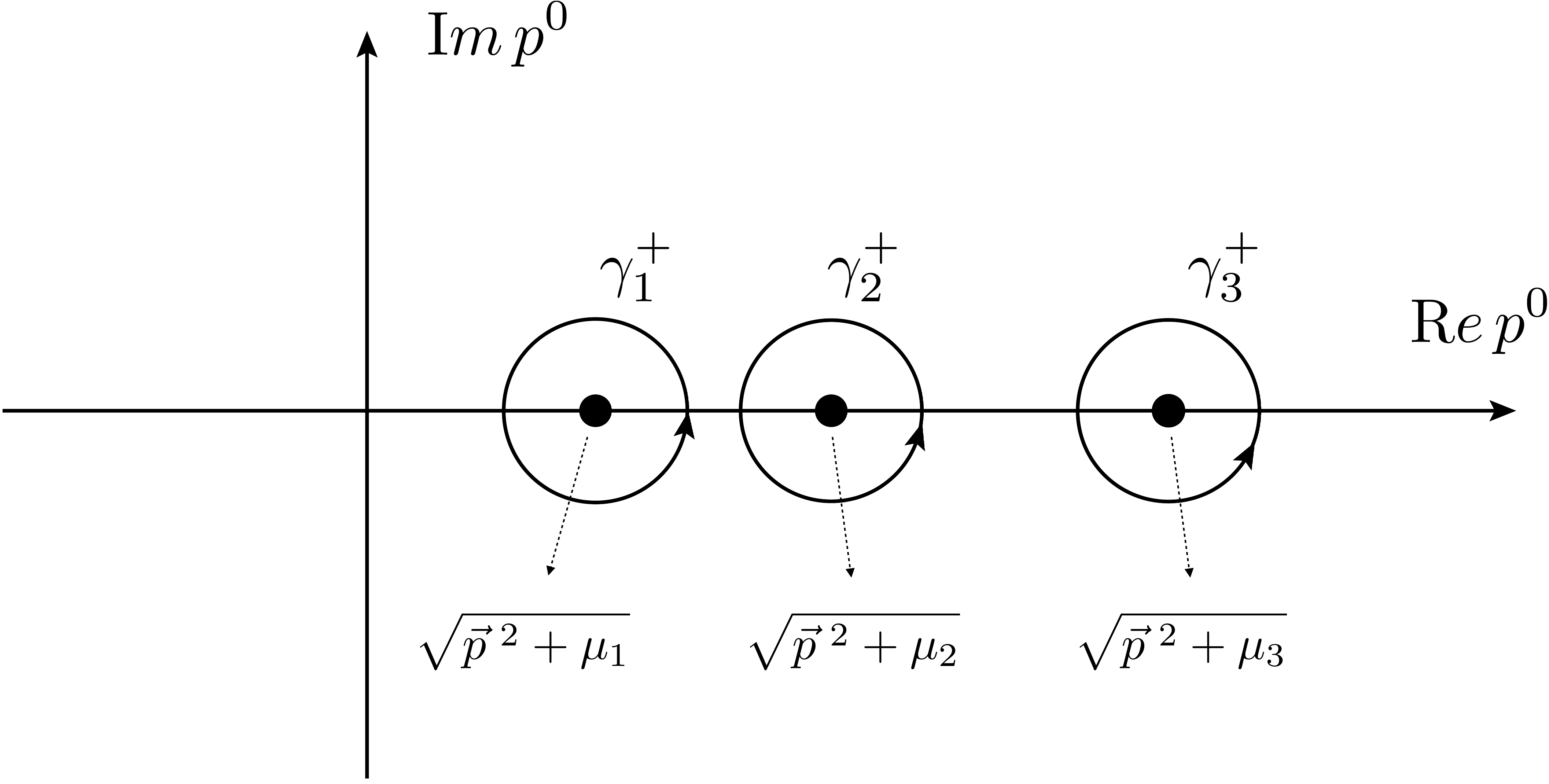}
		\caption{\label{fig2} Integration contours for the positive frequency Wightman function based on the Ostrogradski decomposition \eqref{poly2} of the function $\widetilde{G}(p^2)$.}
	\end{centering}
\end{figure}

At this stage the following remark is in order. On mathematical grounds the decomposition \eqref{poly2} works as long as all roots of the polynomial have order one. On physical grounds there are extra conditions on the roots: comparing eqs.\ \eqref{poly2} and \eqref{genfct} establishes that $\mu_i = m^2$ should be identified with the square of the particle mass. This implies that roots located at the negative real axis correspond to modes with a negative mass squared. In this case the isolated poles at $p_0 = \pm \sqrt{\vec{p}^2 + \mu_i}$ are turned into branch cuts and we will not consider this tachyonic case in the following. Moreover, complex roots always come in pairs $\mu, \bar \mu$. This implies that the positive frequency Wightman function contains unstable modes which grow exponentially in the far past and far future (also see \cite{Aslanbeigi:2014zva} for a detailed discussion of this feature).
On this basis, we restrict ourselves to polynomials $P_n(p^2)$ whose roots
are located at the positive real axis, see Fig.\ \ref{fig2}.

Since the rate function \eqref{parkereq2} is linear in the Wightman function, it is rather straightforward to obtain the detector response function for the case \eqref{poly2}. Following the steps of Sect.\ \ref{sect.23}, we can compute the profile function ${\cal F}(E)$ determining the rate  \eqref{defresf}. Substituting the explicit form of the $A_i$ from \eqref{Aisol} the result reads
\be\label{moddetrate}
{\cal F}  (E) = \frac{1}{c} \, \sum_{i=1}^n \; \left( {\prod_{j \not = i} (\mu_i - \mu_j)} \right)^{-1} \, 
\sqrt{E^2 - \mu_i}\; \theta( E -\sqrt{\mu_i}) \, . 
\ee
The rate function is completely determined by the roots of the polynomial ${\cal P}_n(p^2)$. It receives new contributions once new channels become available, i.e., if the energy gap $E$ crosses a threshold $\mu_i$ where new degrees of freedom enter. Ordering the roots $\mu_i$ by their magnitude, i.e., $\mu_j > \mu_i$ for $j > i$, one sees that the sector with $\mu_j$, $j > i$ does not affect the ``low-energy'' part of the rate function with $E < \mu_i$: the energy gap $E$ of the detector is not large enough to absorb a particle of mass $\sqrt{\mu_j}$, $j > i$. This, in particular, implies that if the polynomial \eqref{poly1} arises from an effective field theory description of a system, there are no corrections to the massless Unruh effect below the first threshold $\mu_2 > 0$, provided that the polynomial ${\cal P}_n$ is properly normalized. The master formula \eqref{moddetrate} then constitutes the main result of this section.

\subsection{Detector rates from the K{\"a}llen-Lehmann representation}
\label{sect.32}
Notably, not all two-point functions proposed in the context of quantum gravity fall in the class where the Ostrogradski-type decomposition is admissible. A prototypical example is provided by Causal Set Theory. Here $\widetilde{G}(p^2)$ interpolates between the standard propagator for a massive scalar field for momenta $p^2$ below the discretization scale and a nonlocal expression without giving rise to additional poles in the complex $p^0$-plane \cite{Carlip:2015mra,Belenchia:2015aia}. In these cases it is still possible to obtain an explicit formula for the profile function ${\cal F}(E)$ based on the K{\"a}llen-Lehmann representation of the two-point function.

The K{\"a}llen-Lehmann representation of the positive frequency Wightman function in position space is given by
\be
G_+(t,\vec{x}) = \int_0^{\infty} dm^2 \, \rho(m^2) \, G_+^{(0)}(t,\vec{x}; m) \, . 
\ee
Here $\rho(m^2)$ denotes a spectral density and $G_+^{(0)}(t,\vec{x}; m)$ is the positive-frequency Wightman function given
in eq.\ \eqref{massiveWightman}. Substituting the K{\"a}llen-Lehmann representation into \eqref{parkereq2} and exchanging the order of integration, the computation of the rate function reduces to the one for the massive scalar field carried out in Sect.\ \ref{sect.23}. The resulting profile function ${\cal F}(E)$, eq.\ \eqref{defresf}, is given by  
\be
\label{spectralF}
\begin{split}
{\cal F}  (E)  
= & \,  \int_0^{E^2} dm^2 \, \rho(m^2) \, \sqrt{E^2 - m^2} \, . 
\end{split}
\ee
Hence the profile function obtained from the K{\"a}llen-Lehmann representation is given by the superposition of contributions with mass $m$ weighted by the spectral density $\rho(m^2)$. Only excitations with mass 
below the energy gap of the detector contribute to the rate function, which is consistent with the expectation that contributions with $m^2>E^2$ will not excite the detector. The result from the Ostrogradski decomposition, eq.\ \eqref{moddetrate}, can then be understood as a special case where $\rho(m^2)$ is given by a sum of $\delta$-distributions located at $m^2 = \mu_i$. 

Dimensional reduction in general seems to be at odd with unitarity.
On a manifold with spectral dimension $d_s$, the asymptotic form of the two-point function
in momentum space is
\be
G(p^2) \sim (p^2)^{d/d_s} \, . 
\ee
Expressing a general two-point function through the K{\"a}llen-Lehmann representation as in the previous section,
we see that, as soon as $d_s<d$, its fall-off properties 
can only be consistent with the $p^{-2}$ behavior of the spectral representation
if we relax the positivity properties of the spectral function $\rho (m^2)$.
This automatically entails the presence of negative-normed states and thus a departure from unitarity.

This signals the fact that these types of higher derivative toy models shouldn't be taken too fundamentally.
It is likely that dimensional reduction, together with (local) Lorentz invariance, signals the presence of
a fundamentally nonlocal theory at small scales. The issue of unitarity for nonlocal theories then is more subtle, see 
 \cite{Eliezer:1989cr} for a more detailed discussion.
The higher-derivative toy models can be considered as approximations to a full nonlocal theory,
in which unitarity is preserved.

\section{Scaling dimensions}
\label{sect.3b}
The two-point function $\widetilde{G}(p^2)$ serves as the essential input
for computing both the spectral dimension $D_s$ seen by a scalar field propagating on the spacetime as well as the rate function of the Unruh detector. Thus, it is conceivable that there is a relation between the rate function of the Unruh detector and the spectral dimension. This section  introduces the definitions needed to make this relation precise.

In the computation of the spectral dimension, $p^2 \equiv (p^0)^2 - \vec{p}^{\,2}$ is analytically continued to Euclidean signature $p^2_E \equiv   (p^0_E)^2 + \vec{p}^{\,2}  > 0$. Subsequently, one introduces a fiducial  diffusion process based on a (modified) diffusion equation
\be\label{diffeq}
\partial_\sigma \, K(x,x^\prime; \sigma) = - F(-\partial^2_E) \, K(x,x^\prime; \sigma) \, , 
\ee
subject to the boundary condition $K(x,x^\prime, 0) = \delta^d(x-x^\prime)$.
Here $\sigma$ is the (external) diffusion time, $K(x,x^\prime; \sigma)$ is the diffusion kernel and $F(-\partial^2_E)$ is determined by the equations of motion of the propagating field. In terms of Fourier-modes $F(p^2_E) = (\widetilde{G}(-p^2_E))^{-1}$. The solution of Eq.\ \eqref{diffeq} is readily obtained in Fourier-space and reads   
\be
K(x,x^\prime; \sigma) = \int \frac{d^dp}{(2\pi)^d} \, e^{i p (x-x^\prime)} \, 
e^{- \sigma F(p^2_E)} \, . 
\ee
The return probability after diffusion time $\sigma$ is given by
\be\label{retprob}
P(\sigma) = \int \frac{d^dp}{(2\pi)^d} \,  
e^{- \sigma F(p^2_E)} \, ,
\ee
and the scale-dependent spectral dimension $D_s(\sigma)$ is defined as
\be
D_s(\sigma) = -2 \frac{d \ln P(\sigma)}{d \ln \sigma} \, . 
\ee
This definition generalizes the standard definition of the spectral dimension $d_s$ which is recovered by evoking the limit of infinitesimal random walks $\sigma \rightarrow 0$. This framework yields the spectral dimension associated with the two-point function $\widetilde{G}(p^2)$ commonly used to assess the dimensionality of spacetime in quantum gravity. 

Analyzing the scaling behavior in \eqref{retprob} one finds that for the case where $F(p^2_E) \propto p^{2 + \eta}_E$ the spectral dimension is given by \cite{Reuter:2011ah}
\be\label{spectraldimension}
D_s = \frac{2d}{2+\eta} \, . 
\ee
The case of a massless scalar field with $\widetilde{G}(p^2) = p^{-2}$ corresponds to $\eta = 0$ and the spectral dimension agrees with the topological dimension $d$ of the spacetime. In case of a multiscale geometry
the scaling law $F(p^2_E) \propto p^{2 + \eta}_E$ is obeyed for a certain interval of momenta only. In this case the spectral dimension will depend on the diffusion time $\sigma$. If the scaling regime extends over a sufficiently large order of magnitudes, $D_s(\sigma)$ will be approximately constant in this regime, realizing a plateau structure. Typically, such plateaus where $D_s(\sigma)$ is approximately constant are connected by short transition regions where $D_s$ changes rather rapidly, see Fig.\ \ref{Fig.dimflow1} for an explicit example illustrating this type of crossover. 

In a similar spirit, one can define the effective dimension of spacetime seen by the Unruh detector. Eq.\ \eqref{rategen} indicates that the profile function for a massless scalar field obeying the Klein-Gordon equation in a $d$-dimensional spacetime scales as
\be
{\cal F}(E) \propto E^{d-3} \, . 
\ee
This motivates defining the effective dimension  seen by the Unruh rate, the Unruh dimension $D_U$, according to
\be\label{UnruhDimension}
D_U(E) \equiv \frac{d \ln {\cal F}(E)}{d \ln E} + 3 \, . 
\ee
For a massless scalar field with $\widetilde{G}(p^2) = p^{-2}$ or a massive
scalar field with energy $E^2 \gg m^2$, $D_U$ is independent of $E$ and coincides with the classical dimension $d$ of the underlying spacetime.
Paralleling the discussion of the spectral dimension, this feature changes, however, if $\widetilde{G}(p^2)$ has a non-trivial momentum profile. The examples presented in Sect.\ \ref{sect.4} indicate that $D_U$ may agree with the spectral dimension in certain cases, but in general the two are different quantities. The Unruh dimension may yield a characterization of quantum spacetimes which is accessible by experiment, at least in principle. 
Note that the dimensions are only well-defined in plateau regions of sufficient extent and have to be taken with caution during crossovers \cite{Reuter:2011ah}.

A direct comparison between $D_U$ and $D_s$ requires an identification of $E$ and the diffusion time $\sigma$. The matching of dimensions in the 
classical case suggests using
\be\label{scalesetting}
\sigma = E^{-2n} \, , 
\ee
where $2n$ is the mass-dimension of $\widetilde{G}(p^2)$. 
We will use this relation in the sequel.

The emission/absorption rates can be related to the density of states of the system
interacting with the detector.
The density of states as a function of momentum can be defined as $\rho(k) = d\Omega(k)/dk$,
where $\Omega(k)$ is the volume of momentum space.
Since the spectral dimension $d_s$ is the Hausdorff dimension of momentum space,
we can assume that $\Omega$ will scale as $\Omega(k) \sim c k^{d_s}$.
Then we see that $\rho(k) \propto k^{d_s-1}$, and a smaller value of $d_s$ entails
a suppression of the density of states.
This in turn will imply a suppression of the various transition rates.
Due to the relation between this density of states and the transition rates,
we expect a relation between the spectral and Unruh dimensions, $D_s$ and $D_U$.
This relation will indeed be made more precise in the next sections.

\section{Unruh rates and dimensional flows}
\label{sect.4}
We illustrate the general formalism devised in Sect.\ \ref{sect.3} by first studying corrections to the Unruh rate arising within quantum gravity inspired multiscale models in Sect.\ \ref{sect.41}. The connection to Kaluza-Klein theories, spectral actions, and Causal Set Theory will be made in Sects.\ \ref{sect.43}, \ref{sect.44}, and \ref{sect.45}, respectively.

\subsection{Dynamical dimensional reduction}
\label{sect.41}
In this subsection we investigate modifications of the Unruh rate arising from a particular class of quantum-gravity inspired two-point functions $\widetilde{G}(p^2)$ typically encountered when discussing the flows of the spectral dimension. 
\subsubsection*{Two-scale models}
The simplest way to obtain a system exhibiting dynamical dimensional reduction is based on a polynomial, eq.\ \eqref{poly1} with $n=2$, containing a single mass scale $m$:
\be\label{ansatz1}
\cP_2(p^2) = - \frac{1}{m^2} \, p^2 \, \left( p^2 - m^2 \right) \, . 
\ee
Here the normalization $c$ has been chosen such that the model gives rise to a canonically normalized two-point function at low energy.
The scaling of this ansatz is given by
\be
\cP_2(p^2) \propto
\left\{	
\begin{array}{ll}
p^2 \, , \qquad & p^2 \ll m^2 \\[1.1ex]
p^4 \, , \qquad & p^2 \gg m^2 \, , 
\end{array}
\right.
\ee
with the crossover occurring at $m^2$. Evaluating \eqref{spectraldimension}, the spectral dimension based on this model interpolates between a classical regime with $D_s = 4$ for long diffusion times and $D_s = 2$ for short diffusion times.
\begin{figure}
	\begin{centering}
		\includegraphics[width = 0.45\textwidth]{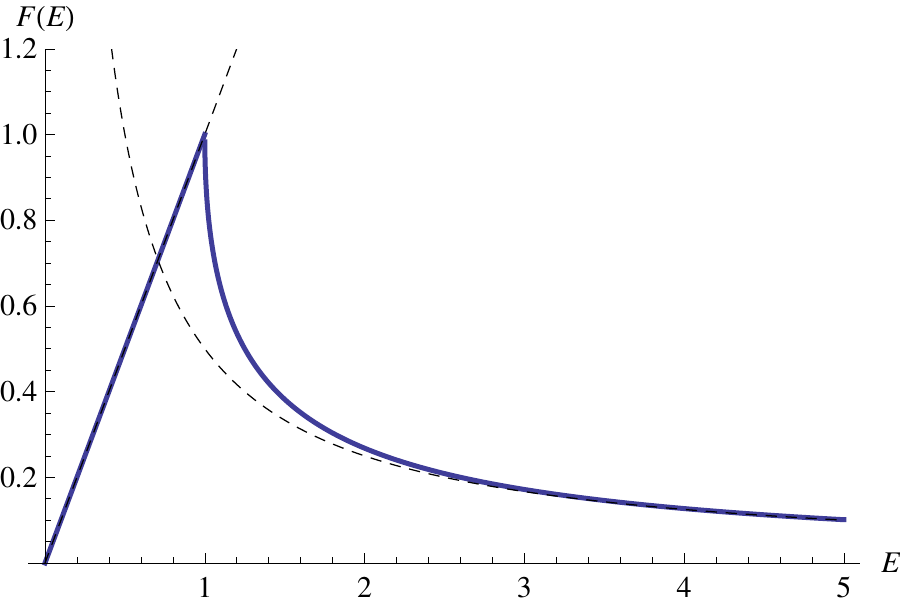} \;
		\includegraphics[width = 0.45\textwidth]{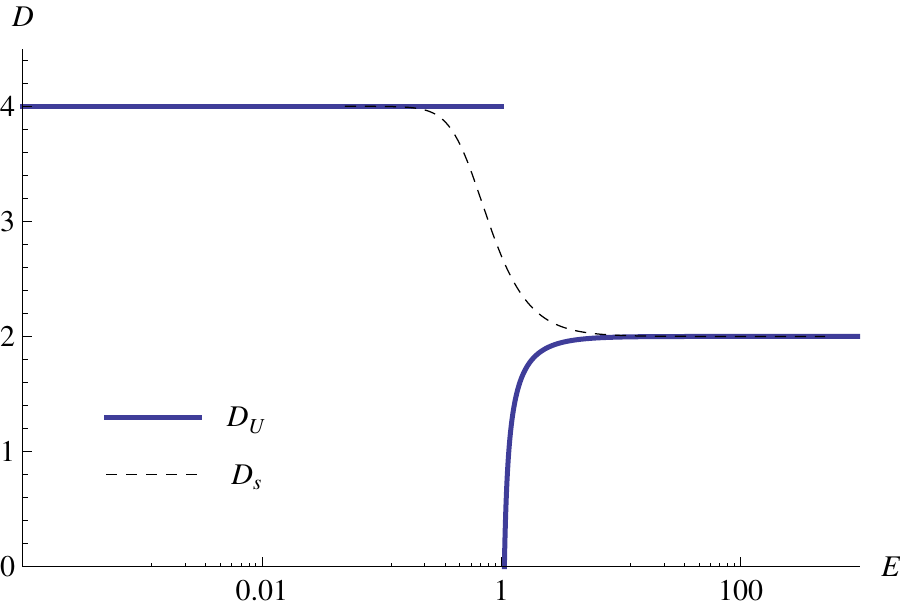}
		\caption{\label{Fig.dimflow1} Profile function $\cF(E)$, eq.\ \eqref{2scalemodel}, for $m=1$ (left panel). The asymptotics given in eq.\ \eqref{2scaleasym} are illustrated by the dashed lines. The right panel shows the dimensions $D_s$ (dashed line) and $D_U$ (solid line) resulting from the two-point function \eqref{2point}.}
	\end{centering}
\end{figure} 

The Ostrogradski decomposition \eqref{poly2} of \eqref{ansatz1} yields
\be\label{2point}
\widetilde{G}(p^2)  = \frac{1}{p^2} - \frac{1}{p^2 - m^2} \, . 
\ee
The master formula \eqref{moddetrate}  gives the following expression for the profile function
\be\label{2scalemodel}
{\cal F}  (E) = E - \sqrt{ E^2 - m^2} \, \theta(E - m) \, . 
\ee
Expanding $\cF$ for small and large $E$ leads to the scaling behavior
\be\label{2scaleasym}
\begin{array}{llcl}
	E < m: \qquad & {\cal F}(E) = E    & \quad \Longleftrightarrow \quad & D_U = 4 \, ,  \\[1.2ex]
	E \gg m: \qquad & {\cal F}(E) = \frac{1}{2E} + \cO(E^{-2})  & \quad \Longleftrightarrow \quad & D_U = 2 \, . 
\end{array}
\ee
This expansion implies that a kinetic term including higher-derivative contributions leads to detector rates which are suppressed at high energies.
In particular, whereas for a massless (free or interacting) scalar field with
a standard kinetic term the prefactor of the rate grows linearly with energy,
the profile function vanishes proportional to  $E^{-1}$ at high energies.
This also entails that the Unruh dimension $D_U$ interpolates between the classical dimension $D_U = 4$ for small energy and $D_U=2$ for $E \gg m$.

For $m=1$ this profile function is shown in the left panel of Fig.\ \ref{Fig.dimflow1}. Despite the inclusion of modes with a wrong sign kinetic term (poltergeists) in \eqref{2point} the Unruh rate is positive definite, indicating that the model is stable in this respect. The right panel of Fig.\ \ref{Fig.dimflow1} shows the spectral dimension (dashed line) and effective dimension seen by the Unruh effect (solid line) where the construction of the spectral dimension is based on the identification \eqref{scalesetting}. Both dimensions interpolate between $D = 4$ for $E < m$ and $D = 2$ for $E \gg m$. $D_U$ displays a discontinuity at $E^2 = m^2$ which can be tracked back to the derivative of the square-root becoming singular at this point.

\subsubsection*{Multi-scale models}
At this stage it is instructive to consider a multiscale model which may exhibit more than two scaling regions. The simplest model of this form is build from a third order polynomial $\cP_3(p^2)$ with vanishing mass $m_1 = 0$
\be\label{ansatz2}
\cP_3(p^2) = \frac{1}{m_2^2 \, m_3^2} \, p^2 \, (p^2-m_2^2) \, (p^2 - m_3^2) \, , \qquad m_3 > m_2 \, . 
\ee
Provided that $m_3 \gg m_2$ this ansatz exhibits three scaling regimes 
\be\label{3scalemod}
\cP_3(p^2) \propto
\left\{	
\begin{array}{lll}
	p^2 \, , \qquad & p^2 \ll m^2_2  \, , & D_s = 4\\[1.1ex]
	p^4 \, , \qquad & m_2^2 \ll p^2 \ll m^2_3  \, , \qquad & D_s = 2 \\[1.2ex]
	p^6 \, , \qquad & m_2^3 \gg p^2  \, , & D_s = \frac{4}{3} \, ,  \\[1.1ex]
\end{array}
\right.
\ee
where the spectral dimension has been determined by evaluating \eqref{spectraldimension}.

\begin{figure}[t]
	\begin{centering}
		\includegraphics[width = 0.45\textwidth]{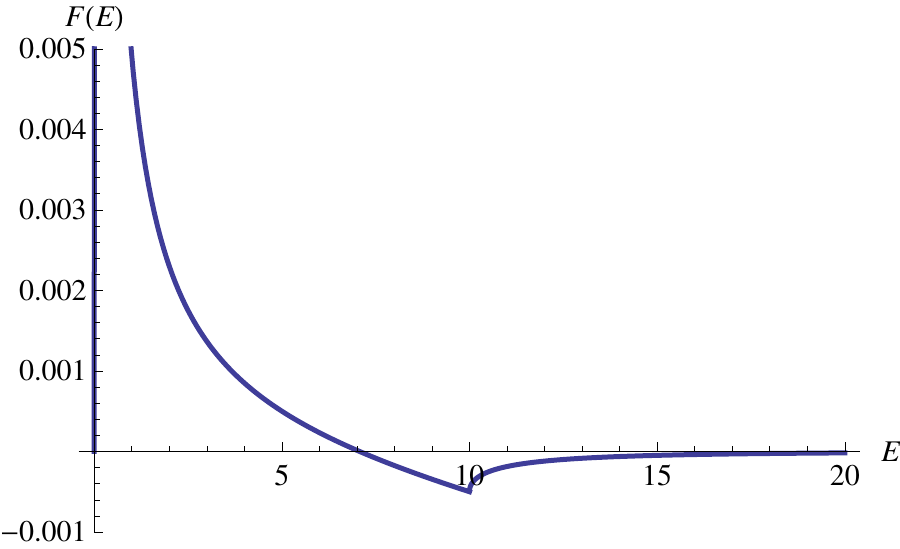} \;
		\includegraphics[width = 0.45\textwidth]{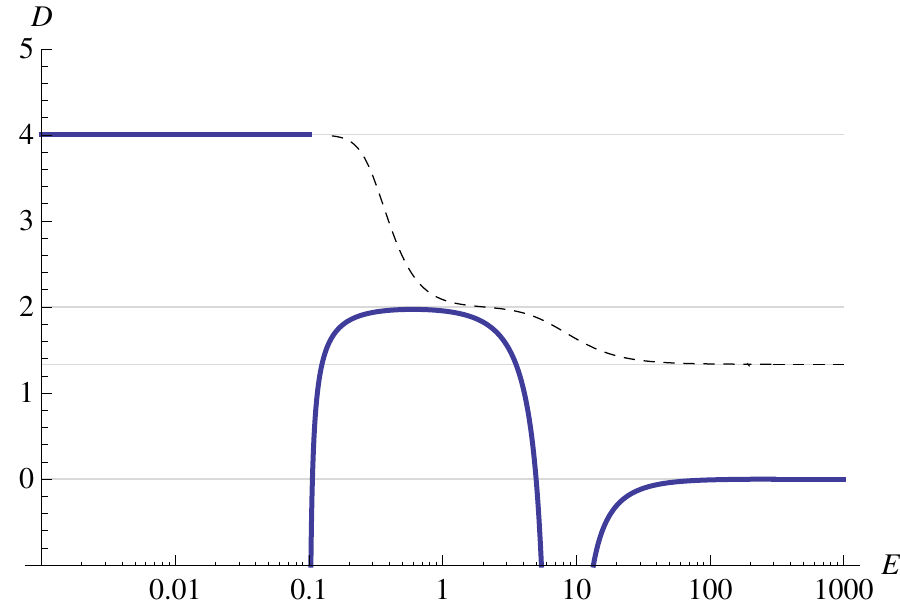}
		\caption{\label{Fig.dimflow2} Illustration of the Unruh effect in a $n=3$ multiscale model with $m_1 = 0$, $m_2 = 0.1$ and $m_3 = 10$. The resulting profile function $\cF(E)$ is shown in the left panel while $D_U$ and $D_s$ are displayed in the right panel. The horizontal gray lines indicate the plateau values of the dimensions at $4,2,4/3$ and $0$. Notably, $D_U$ and $D_s$ exhibit different asymptotics for $E \gg m_3$.}
	\end{centering}
\end{figure} 
Performing the Ostrogradski decomposition for $\cP_3(p^2)$ gives
\be
\widetilde{G}(p^2) = \frac{1}{p^2} - \frac{m_3^2}{m_3^2 - m_2^2} \, \frac{1}{p^2 - m_2^2} + \frac{m_2^2}{m_3^2 - m_2^2} \, \frac{1}{p^2 - m_3^2} \, . 
\ee
The resulting profile function then reads
\be
\cF(E) = E - \frac{m_3^2}{m_3^2 - m_2^2} \, \sqrt{E^2 -m_2^2} \, \, \theta(E-m_2) + \frac{m_2^2}{m_3^2 - m_2^2} \, \, \sqrt{E^2 -m_3^2} \, \, \theta(E-m_3) \, . 
\ee
Expanding $\cF$ for small and large $E$ leads to the scaling behavior
\be\label{3scaleasym}
\begin{array}{llcl}
	E < m_2: \qquad & {\cal F}(E) = E    & \quad \Longleftrightarrow \quad & D_U = 4 \, ,  \\[1.2ex]
	E \gg m_3: \qquad & {\cal F}(E) = - \frac{m_2^2 \, m_3^2}{8E^3} + \cO(E^{-4})  & \quad \Longleftrightarrow \quad & D_U = 0 \, . 
\end{array}
\ee

At this stage two remarks are in order. In contrast to the two-scale model, the $n = 3$ case exhibits regions where the profile function $\cF(E)$ actually becomes negative. This is illustrated in the example shown in Fig.\ \ref{Fig.dimflow2}. The form where $\lim_{E \rightarrow \infty}F(E) \rightarrow 0$ from below then indicates that this feature holds for all values $m_2$ and $m_3$. Thus the Unruh rate exhibits an instability for a generic $n=3$ model.

Furthermore, the spectral and Unruh dimensions shown in the right panel of Fig.\ \ref{Fig.dimflow2} show that, contrary to the two-scale model, the asymptotics for $D_U$ and $D_s$ do not agree for $E \gg m_3^2$. In the general case, this may be understood as follows. Considering the general expression \eqref{moddetrate} for $m_1 = 0$, $D_U$ is given by the classical dimension as long as $E < m_2$. Each additional term in the sum creates a new scaling region where $D_U$ decreases by two compared to its previous value. In contrast the pattern for the spectral dimension follows from \eqref{spectraldimension}. Combining these relations allows to express the effective dimension seen by the Unruh effect in terms of the spectral dimension
\be\label{dimrel}
D_U = 6 - \frac{8}{D_s} \, . 
\ee
Thus, while there is a clear relation between $D_U$ and $D_s$, the effective dimensions seen by a random walk and the Unruh effect generically do not coincide within the class of multiscale models studied here.

\subsubsection*{Logarithmic correlation functions}
An interesting model which does not fall into the class of multiscale models where the Ostrogradski decomposition can be applied arises from
\be\label{p4model}
\widetilde{G}(p^2) = p^{-4} \, . 
\ee
This is the typical fall-off behavior of correlation functions in quantum gravity models which lead to $D_s = 2$ in the ultraviolet. 
In this case the positive-frequency Wightman function is
\be
G_+(\vec{x},t) = -i \int \frac{d^3 k}{(2\pi)^3} \oint_{\gamma_+} \frac{dk^0}{2\pi} \frac{e^{i\vec{k}\cdot\vec{x} - ik^0 t}}{(k^0 + | \vec{k} |)^2(k^0 - | \vec{k} |)^2} \, . 
\ee
Picking up the double pole at $k^0=|\vec{k}|$, and setting $\vec{x}=0$ before carrying out the angular momentum integral, one obtains
\be\label{I12}
G_+(\vec{x},t) = -4\pi \int_0^{\infty}  \frac{dk}{(2\pi)^3} \, k^2 \, \left[ \frac{2}{(2k)^3} + \frac{it}{(2k)^2} \right] e^{-ik(t- i \epsilon)} = I_1 + I_2 \, . 
\ee
The second integral is simply
\be
I_2 = -\frac{1}{8\pi^2} \, . 
\ee
The first integral can be written as a regularized Laplace transform and gives
\be
I_1 = \lim_{\epsilon\to 0^+} \, \lim_{\tilde{\epsilon}\to 0^+} \,
\Gamma(\tilde{\epsilon}) \left( \epsilon +it \right)^{-\tilde{\epsilon}}
= \frac{1}{8\pi^2}\left(\log t + \mbox{const} \right) \, . 
\ee
Thus the resulting positive frequency Wightman function has a logarithmic dependence on the proper distance. Restoring Lorentz invariance, we get
\be
G_+(\vec{x},t) = \frac{1}{8\pi^2} \left[ \log \left(\sqrt{\left( t - t^{\prime} -i\epsilon \right)^2 - \left( \vec{x} - \vec{x}^{\prime} \right)^2 } \right) + \mbox{const} \right] \, . 
\ee

Substituting the Wightman function into the formula for the Unruh rate, eq.\ \eqref{parkereq2}, yields
\be
\dot{F} (E) = \frac{1}{8\pi^2} \int_{-\infty}^{\infty} d\tau e^{i E \tau} \,
\left[ \log \left( \frac{2\sinh(\frac{a\tau}{2})}{a\tau} \right) + \mbox{const} \right] \, . 
\ee
The constant terms give rise to terms proportional to $\delta(E)$, indicating an infrared instability of the setup. Since the propagator \eqref{p4model} is thought of describing the asymptotic behavior of the system at high energies we will ignore these terms in the following. Since the argument of the logarithm is an even function in $\tau$ the integral can be expressed as a (regularized) Fourier cosine transform

\be
\dot{F} (E)  = \lim_{\epsilon\to 0^+}\frac{1}{2a\pi^2} \int_{0}^{\infty} dx e^{-\epsilon x} \log \left( \frac{\sinh(x)}{x} \right) \, 
\cos(\omega x) \, . 
\ee
written in terms of the new variables $x=a\tau/2$ and $\omega=2E/a$.
This integral can now be written as $I=I_1 - I_2$, where
\be
\begin{split}
I_1 = & \, \lim_{\epsilon\to 0^+} \;  \frac{1}{2a\pi^2} \frac{d}{d\alpha} \left. \int_{0}^{\infty} dx e^{-\epsilon x} 
\left( \sinh(x) \right)^{\alpha} \cos(\omega x) \right|_{\alpha\to 0}
= - \frac{\pi \coth(\frac{\pi\omega}{2})}{2\omega} \, , \\
I_2 = & \, \lim_{\epsilon\to 0^+} \; \frac{1}{2a\pi^2} \frac{d}{d\alpha} \left. \int_{0}^{\infty} dx e^{-\epsilon x} 
x^{\alpha} \cos(\omega x) \right|_{\alpha\to 0}
= - \frac{\pi}{2\omega} \, .
\end{split} 
\ee
Combining the two contributions, the resulting detector rate is given by
\be
\dot{F} (E) = \frac{1}{4\pi E} \, \frac{1}{1-e^{\frac{2\pi E}{a}}} \, ,
\ee
implying that the profile function resulting from a $p^{-4}$ propagator
is given by
\be
{\cal F}(E)= \frac{1}{2E} \qquad \, \Longleftrightarrow \qquad  D_U = 2 \, .   
\ee
This is precisely the asymptotic behavior \eqref{2scaleasym} found in the two-scale model in the limit $E \gg m$. Thus the direct computation of the detector rate in the $p^4$-case confirms the drop of the Unruh rate at high energies and constitutes an independent verification of the rate function found in the two-scale case.

\subsection{Kaluza-Klein theories}
\label{sect.43}
A scenario where the dimensional reduction occurs when going 
towards the \emph{infrared} is provided by Kaluza-Klein theories.\footnote{A related discussion of the Unruh detector in Kaluza-Klein theories appeared in Ref.\ \cite{Chiou:2016exd} during the final stage of preparing the manuscript.}
In this case the (classical) spacetime is assumed to possess four non-compact and a number of compact spatial dimensions whose typical extension is given by the compactification scale $R$. At length scales $l \gg R$ the effect of the extra-dimensions is invisible and physics is effectively four-dimensional. We demonstrate that also in this situation the dimensional reduction entails a suppression in the Unruh effect. In the case of Kaluza-Klein theories where the number of effective dimensions increases when going to high energies this implies that the detector rates for energies above the inverse compactification scale are actually \emph{enhanced} as compared to the four-dimensional rate.

\begin{figure}
	\begin{centering}
		\includegraphics[width=0.45\textwidth]{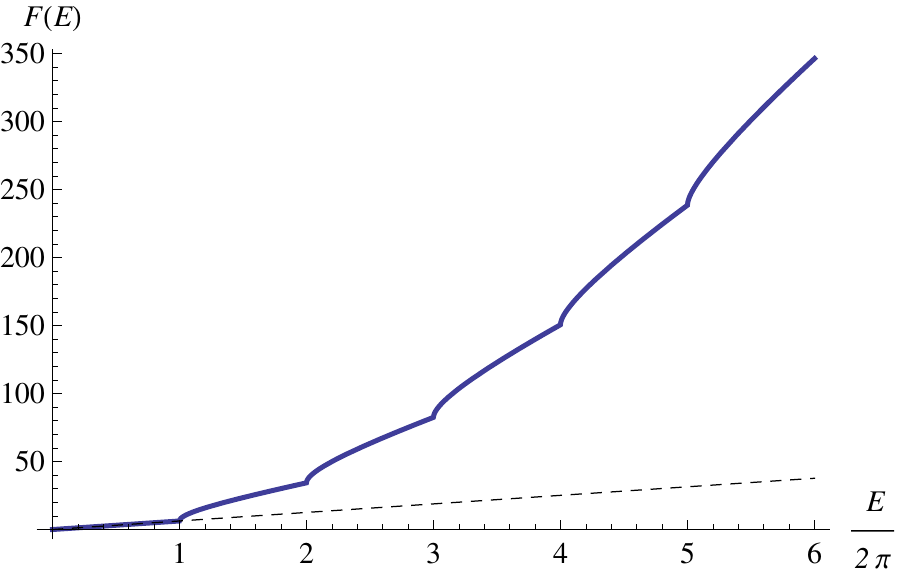} \, 
		\includegraphics[width=0.45\textwidth]{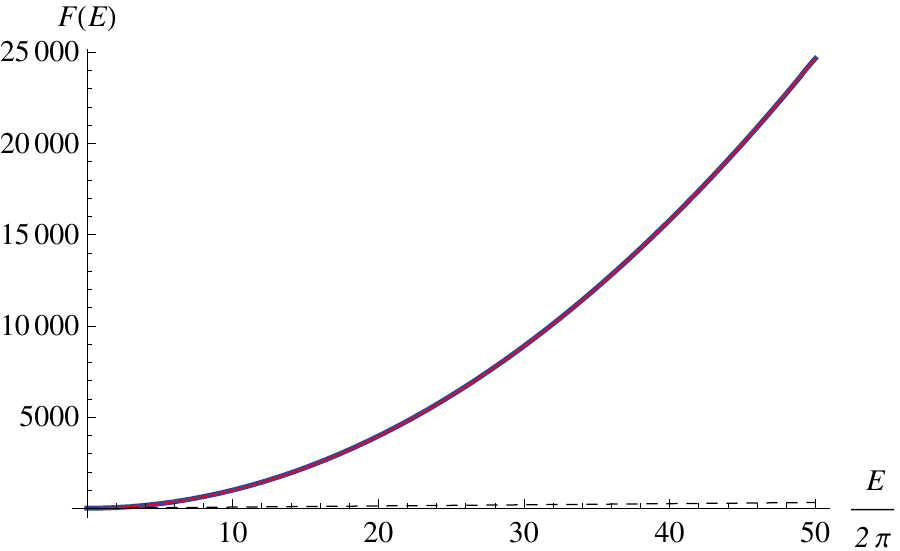}
		\caption{\label{FigKKprofile} Profile function ${\cal F}(E)$ for a $5$-dimensional Kaluza-Klein theory \eqref{kkprofile} with $R = 1/(2\pi)$ (blue, solid line). For guidance the lines ${\cal F}(E) = E$ (black, dashed line)  and ${\cal F}(E) = E^2/4$ (red line, right diagram) have been included.
			For $E < R^{-1}$ the profile function is linear in $E$, while for $E \gg R^{-1}$ it increases proportional to  $E^2$.
		}
	\end{centering}
\end{figure}

For concreteness we will focus on the case of a five-dimensional spacetime ${\mathbb R}^4 \times S^1_R$ where the extra dimension is given by a compact circle of radius $R$. A scalar field $\phi$ living on this spacetime has a Fourier-expansion in the circle coordinate $x_5$
\be
\phi(x,x_5) = \sum_{n=-\infty}^{+\infty} \phi_n(x) \, e^{i\frac{n}{R}x_5} \, , \qquad x_5 \in [0, 2 \pi R[ \, . 
\ee
The Fourier coefficients $\phi_n(x)$ depend on the coordinates on ${\mathbb R}^4$ and are called Kaluza-Klein modes. For a real scalar field $\phi$ they obey the reality condition $\phi_{-n} = \phi_n^*$. Substituting this mode expansion into the action of a free scalar field in five dimensions yields
\be
\int d^5x \, \tfrac{1}{2} \, \left[ (\partial_{\mu}\phi)^2 - (\partial_5 \phi)^2 \right] =
 \, 2 \pi R \int d^4x \sum_{n=-\infty}^{+\infty} \tfrac{1}{2} \, \left[ |\partial_{\mu}\phi_n |^2 - \frac{n^2}{R^2} |\phi_n |^2 \right] \, . 
\ee
Each Kaluza-Klein mode $\phi_n$ has a two-point function of a scalar field with mass $m_n = n/R$. Taking into account the entire tower of modes, the resulting function $\widetilde{G}(p^2)$ is given by
\be
\widetilde{G}(p^2) = \frac{1}{2 \pi R} \, \sum_{n=-\infty}^{\infty} \, \left( p^2 - \frac{n^2}{R^2}\right)^{-1} \, . 
\ee
Applying the master formula \eqref{moddetrate} to this case then yields the profile function
\be\label{kkprofile}
{\cal F}  (E) = \frac{1}{2\pi R} \, \left( E + 2 \sum_{n=1}^\infty \sqrt{ E^2 - (n/R)^2} \; \theta( E - n/R) \right) \, . 
\ee
The shape of this profile function is illustrated in Fig.\ \ref{FigKKprofile}.
In contrast to the case of a dynamical dimensional reduction at high energies, all Kaluza-Klein modes contribute to the profile function with the same sign. This leads to an effective enhancement of the profile function for $E > R^{-1}$.
Explicitly,
\be
\begin{array}{llcl}
E < 1/R: \qquad & {\cal F}(E) \propto E    & \quad \Longleftrightarrow \quad & D_U = 4 \, ,  \\[1.2ex]
E \gg 1/R: \qquad & {\cal F}(E) \propto E^2  & \quad \Longleftrightarrow \quad & D_U = 5 \, . 
\end{array}
\ee
 The profile function \eqref{kkprofile} interpolates between these two behaviors. Thus also the presence of extra dimensions leaves its imprint on the Unruh rate, adapting the scaling law of the profile function once the energy $E$ exceeds the inverse compactification scale.

\subsection{Spectral actions}
\label{sect.44}
A framework which naturally gives rise to two-point functions
$\widetilde{G}(p^2)$ with the properties discussed above are spectral
actions. The basic idea is that the action describing the dynamics of
the theory is generated by the trace of a suitable differential operator, typically the Dirac operator $\cD$
\begin{equation}
S_{\chi,\Lambda}=\Tr \bigl[ \chi(\cD^2/\Lambda^2) \bigr] \, . 
\label{SpAct1}
\end{equation}
Here $\chi$ is a positive function and $\Lambda$  sets the typical scale of the theory. Spectral actions provide the core ingredient for setting up a geometrical formulation of the standard model of particle physics based on almost-commutative geometries \cite{Chamseddine:1996zu,Chamseddine:1996rw}, also see \cite{MarcolliBook,vandenDungen:2012ky,vanSuijlekom:2015iaa} for reviews. Here we focus on the case where $\cD^2$ is given in terms of the Laplace operator on flat Euclidean space supplemented by an endomorphism including a real scalar field $\phi$:\footnote{The spectral dimension arising in this situation has recently been studied in \cite{Alkofer:2014raa}, also see \cite{Kurkov:2013kfa} for a related discussion.}
\be\label{lapop}
\cD^2 = - \left( \nabla^2 \mathbb{1} + E \right) \, , \quad E = - i \gamma^\mu \gamma_5 \partial_\mu \, \phi - \phi^2 \, .  
\ee
The definition of the model is then completed by specifying the function $\chi$.
%

\subsubsection*{Nonlocal analytic models} 
We first discuss the case where $\chi (z) = e^{-z}$. In this case the
spectral action \eqref{SpAct1} coincides with the heat-trace of the Laplace-type operator \eqref{lapop} which is a well-studied mathematical object, see e.g., \cite{Vassilevich:2003xt,Barvinsky:1987uw,Barvinsky:1990up,Avramidi:2000bm,Iochum:2011yq,Codello:2012kq}. In particular the two-point function of the model is given by 
\be\label{spectral2}
S^{(2,\phi)}_{\chi,\Lambda}  =  \frac  {\Lambda^2}{(4\pi)^2} \int d^4x \left[
\, \phi \, F_{0} (-\partial^2_E / \Lambda ^2) \, \phi \, 
\right] .
\ee
The structure function $F_{0}$ is obtained from the heat-kernel result for the endomorphism $E$ and reads \cite{Kurkov:2013kfa}
\begin{eqnarray}
F_{0} \left (z\right) =     
2 \, z \, h\left(z\right) - 4  \, ,
\label{F0} 
\end{eqnarray}
with
\begin{equation}
h(z)= \int_0^1 d\alpha \,e^{-\alpha (1-\alpha) z} .
\end{equation}
The function $h(z)$ is an entire analytic function which is nowhere vanishing in the complex plane. The momentum-dependent two-point function for this model is then obtained by analytically continuing \eqref{spectral2} to Lorentzian signature
\be\label{ss1}
\widetilde{G}(p^2) = - \frac{8 \pi^2}{\Lambda^2} \, \frac{1}{F_0(-p^2/\Lambda^2)} \, , 
\ee
where $p^2$ is the Lorentzian momentum four-vector.

A careful study of the two-point function \eqref{ss1} reveals several remarkable features. First, the model naturally gives rise to a Higgs mechanism for $\phi$. The propagator exhibits a pole at $p^2 \simeq -3.41 \Lambda^2$ indicating that the expansion of $\phi$ around vanishing field value corresponds to expanding at an unstable point in the potential. Restoring the $\phi^4$ term\footnote{For a discussion of the Higgs mechanism in almost-commutative geometry see Sect.~11.3.2 of \cite{vanSuijlekom:2015iaa}.}
leads to a scalar potential
\begin{equation}
V(\phi) = -\mu_H^2 \phi^2 + \lambda \phi^4 + \ldots  \, , 
\end{equation}
with $\mu_H^2 = 2 \Lambda^2$. Neglecting the higher-order terms, the potential gives a non-vanishing vacuum expectation value $\langle \phi \rangle = \pm \frac{\mu_H}{\sqrt{2\lambda}}$. Expanding the field around this minimum leads to a potential for the fluctuation field $\tilde \phi$
\be
V(\tilde \phi) = 2 \, \mu_H^2 \, \tilde \phi^2 + \ldots \, , 
\ee
Thus, when expanded around the minimum of the scalar potential, the structure function entering into \eqref{ss1} should be given by
\be\label{Fhiggsed}
F_H(z) = 2 \, z \, h(z) + 8 \, . 
\ee
$F_H(z)$ has a single real root located at $p^2 \simeq 2.56 \Lambda^2$. This root corresponds to a positive mass pole in \eqref{ss1}. In addition there are complex roots located, e.g., at
\be
p^2 = - \left( 1.32 \pm 21.98 i \right) \Lambda^2 \, . 
\ee
These roots can be traced back to the mass-term contribution in $F_0$ or $F_H$ and are absent if one considers the $z h(z)$ part only. The presence of complex roots signals that the Wightman function contains modes which increase exponentially for large times. These modes introduce an instability in the Unruh effect, which we will not investigate further. It would be very interesting to see if there are functions $\chi$ which give rise to a nonlocal theory avoiding this instability.

\subsubsection*{Ostrogradski-type models}
By making a suitable choice for the function $\chi$ one can also generate spectral actions which are local in the sense that the (inverse) two-point function is given by a finite polynomial in $p^2$.\footnote{This is closely related to the zeta-function spectral action proposed in \cite{Kurkov:2014twa}.}
The simplest choice, leading to a two-scale model, uses
\be\label{chi1}
\chi(z) = (a + z) \, \theta (1-z) \, \, , \quad a > 0 \, .
\ee
Replacing the polynomial multiplying the stepfunction by a polynomial of order $n$ leads to a multiscale model whose inverse propagator is given by a polynomial of order $n$ in $p^2$.

The spectral action for these cases can be found explicitly by combining the early-time expansion of the heat-kernel in $s \equiv \Lambda^{-2}$  
\be
\begin{split}
F_H = & \, \frac{1}{(4\pi)^2} \frac{1}{s} \, \sum_{m=0}^\infty \, a_m \, (p_E^2 \, s)^m \, , \\
= & \frac{1}{(4\pi)^2} \frac{1}{s} \left( 8 + 2 \, s \, p_E^2  - \frac{1}{3} \left( s \, p_E^2 \right)^2 + \ldots \right)
\end{split}
\ee
with standard Mellin transform techniques \cite{Codello:2008vh}
\be\label{spectralas}
S^{(2,\phi)}_{\chi,\Lambda}  =  \frac  {1}{(4\pi)^2} \int \frac{d^4p}{(2\pi)^4} 
\, \phi \, \left[ \sum_{m=0} \, Q_{m+1}[\chi] \, a_m \,   (p_E^2)^m \,\right] \, \phi \,  .
\ee
The moments $Q_n$ depend on the function $\chi$ and, for $n \in {\mathbb Z}$ are given by
\be
\begin{array}{ll}
Q_n[\chi] =  \frac{1}{\Gamma(n)} \int^\infty_0 dz \, z^{n-1} \, \chi(z) \, , \qquad & n > 0 \, , \\[1.3ex]
Q_{-n}[\chi] =  (-1)^n \, \chi^{(n)}(0) \, , \qquad & n \ge 0 \, .
\end{array}
\ee
\begin{figure}[t]
	\begin{centering}
		\includegraphics[width=0.45\textwidth]{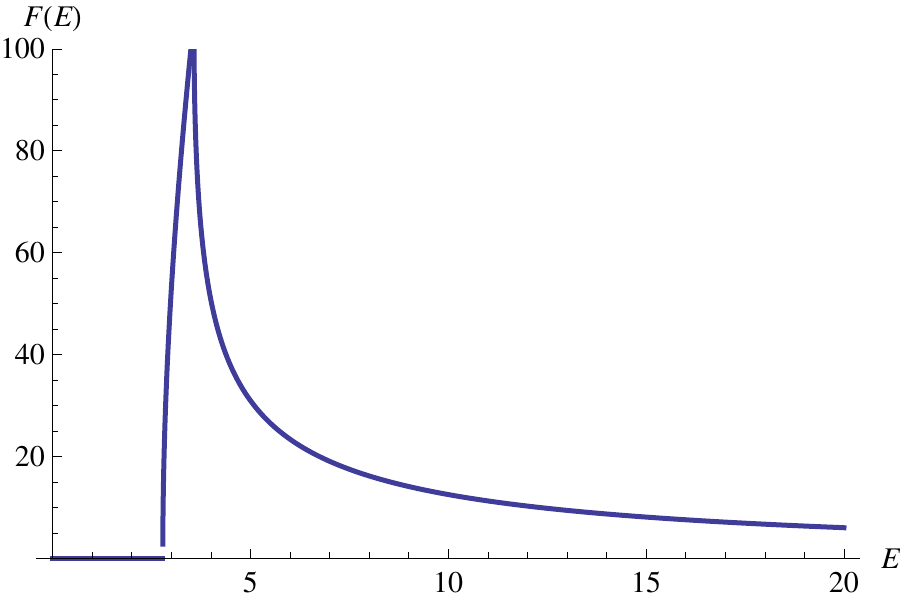}
		\caption{\label{spectralrate} Profile function \eqref{spectralprofile} for $a=3.2$.
		}
	\end{centering}
\end{figure}
For the ansatz \eqref{chi1} the moments are
\be
Q_1[\chi] = a + \frac{1}{2} , \quad Q_0[\chi] = a, \quad Q_{-1}[\chi] = -1 , \quad 
Q_{-2} =Q_{-3} = \ldots =0.
\label{Qchi1}
\ee
Converting to Lorentzian signature, the inverse two-point function based on the expansion of $F_H$, eq.\ \eqref{Fhiggsed} is
\be\label{p3poly}
\cP_2(p^2) = - \frac{1}{8\pi^2} \left( 8a+4 - 2 a p^2 + \frac{1}{3} p^4 \right) \, . 
\ee
The two roots of the system are located at
\be
\mu_{1,2} = 3 a \mp \sqrt{ 9 a^2 - 24 a - 12 } \, . 
\ee
Provided that $2(2 + \sqrt{7})/3 < a < (3 + \sqrt{15})/2$, both roots are on the positive real axis. Thus the model falls into the class discussed in Sect.\ \ref{sect.41}. The profile function is readily obtained by applying the Ostrogradski decomposition to \eqref{p3poly}
\be\label{spectralprofile}
{\cal F}  (E) = \frac{24 \pi^2}{\mu_2 - \mu_1} \left( \sqrt{E^2 -\mu_1} \; \theta( E -\sqrt{\mu_1}) - \sqrt{E^2 - \mu_2} \; \theta( E - \sqrt{\mu_2}) \right) \, . 
\ee
The behavior of this profile function is illustrated in Fig.\ \ref{spectralrate}.

For $E^2 < \mu_1$ the profile function vanishes, indicating that the energy gap is too small for the detector to interact with the two massive fields. For $7.77 < E^2 < 12.77$ the profile corresponds to the standard Unruh rate for a field with mass $m^2 = 7.77$. Once $E^2$ crosses the threshold at $12.77$ the profile function decreases and falls of asymptotically as $E^{-1}$ for high energies. Thus spectral actions may give rise to similar profile functions as the multiscale models discussed at the beginning of this section.
%

\subsection{Causal Set inspired theories}
\label{sect.45}

A second framework which naturally gives rise to corrections to the Unruh effect are the nonlocal two-point functions emerging in the context of Causal Set Theory. In this case the two-point functions extrapolate between a classical massless or massive propagator at energy scales well below the discretization scale and a discrete D'Alembertian naturally associated with the Causal Set at high energies \cite{Aslanbeigi:2014zva,liberati}. 
In this section we will derive the resulting Unruh signature arising from this setting as well as from Causal Set inspired toy models.

\subsubsection*{Rate suppression in the full theory}

The explicit form of the two point function reads\footnote{We set everywhere the sprinkling density $\rho$ to one.}
\begin{equation}\label{2pointcausal}
G_+(x^2)= - \frac{i}{2\pi^3}\int_0^\infty d\xi \xi^2 \frac{K_1(i\sqrt{x^2}\xi)}{\sqrt{x^2}\,\xi \, g(\xi^2)} \, , 
\end{equation}
where $\xi$ is a momentum and
\begin{equation}
g(\xi^2)=a+4\pi \xi^{-1}\sum_{n=0}^{3}\frac{b_n}{n!}C^n\int_0^\infty s^{4(n+1/2)}e^{-Cs}K_1(\xi s)ds \, .
\end{equation}
The parameters are determined based on the analytic properties of the two-point function and given by $a=-\frac{4}{\sqrt{6}}$, $b_0=\frac{4}{\sqrt{6}}$, $b_1=-\frac{36}{\sqrt{6}}$, $b_2=\frac{64}{\sqrt{6}}$, $b_3=-\frac{32}{\sqrt{6}}$, $C=\frac{\pi}{24}$. 
The asymptotics of $g(\xi^2)$ has been determined in \cite{Aslanbeigi:2014zva}
 \begin{equation}\label{asymptotics}
 \begin{split}
 \lim_{\xi^2 \rightarrow 0} \, \frac{1}{g(\xi^2)}= & \, -\frac{1}{\xi^2}+\cdots \, ,  \\ 
 \lim_{\xi^2 \rightarrow \infty} \, \frac{1}{g(\xi^2)}= & \, -\frac{2\sqrt{6}\pi}{\xi^4}+\cdots \, .
 \end{split} 
 \end{equation}
We thus see that at high energies the two-point function has a characteristic $p^{-4}$ behavior.
The profile function will then asymptotically match the result we already derived in Sect.\ 5.1,
for the logarithmic case, displaying a $1/E$ fall off.

Using the two-point function as above, the equation for the detector rate gives
\be
\dot{F}=\frac{-i}{2\pi^3} \int_0^\infty \frac{d\xi \xi^2}{g(\xi^2)}
\int_{-\infty}^{\infty}d\tau e^{-i E\tau}
\left( \frac{K_1(\frac{2i\xi}{a}\sinh{(\frac{a}{2}(\tau-i\epsilon))})}{\frac{2\xi}{a}\sinh{(\frac{a}{2}(\tau-i\epsilon))}}
- \frac{K_1((\tau-i\epsilon)\xi)}{(\tau-i\epsilon)\xi}\right) \, , 
\ee
from which we arrive at the profile function
\be
{\cal F}(E) =-\frac{2}{\pi} \int_0^E d\xi\xi\frac{\sqrt{E^2-\xi^2}}{g(\xi^2)} \, . 
\ee
In principle this relation gives the exact form of the profile function in Causal Set Theory. Its evaluation requires the full form of $g(\xi^2)$ and cannot be based on the asymptotic expansions \eqref{asymptotics} alone. Performing the resulting integral numerically is beyond the scope of the present work. Instead we will focus on a simplified model which allows for an analytic treatment.

\subsubsection*{A consistent toy model}

The central properties of the two-point correlation function for Causal Sets \eqref{2pointcausal} are captured by the combination of a massless pole at zero mass combined with a continuum of states with density $\rho(m^2)$ \cite{Saravani:2015rva}.  
The resulting positive frequency Wightman function is then given by the sum of the massless one, denoted by $G_+^{(0)}$ and an integral over the continuum of states
\be\label{wfcaus2}
G_+(t, \vec{x})=G_+^{(0)}(t, \vec{x}; m=0) + \int_0^{\infty}dm^2 \rho(m^2) \, G_+^{(0)}(t, \vec{x}; m) \, , 
\ee
where $G_+^{(0)}(t, \vec{x}; m)$ denotes the Wightman function for a scalar of mass $m$. Inspired by \cite{Saravani:2015rva} it is conceivable that all relevant physics of the Causal Set construction is retained by approximating the density of states by 
\be
\label{rho}
\rho(m^2) = e^{-\alpha m^2}\sum_{n=0}^N b_n \, m^{2n} \, . 
\ee
Here $\alpha$ is a parameter of order one, $b_0$ is related to the nonlocality scale, and the remaining $b_n$'s are free parameters.

As a consistency requirement, the simplified model should recover the massless theory in the infrared limit. This is ensured by requiring that the continuum contribution to \eqref{wfcaus2} vanishes in the limit where the geodesic distance $Z \equiv (t-t^\prime)^2 - (\vec x - \vec x^\prime)^2$ goes to infinity. Substituting \eqref{rho} into \eqref{wfcaus2} this condition entails
\be
\label{IRcondition}
\lim_{Z \rightarrow \infty} \; \sum_{n=0}^N \, b_n \, \int_0^\infty dm \, e^{-\alpha m^2} \, m^{2n+2} \, \frac{K_1(i m \sqrt{Z})}{\sqrt{Z}} = 0 \, . 
\ee
Applying the expansion of $K_1(x)$ for large argument the resulting integral reduces to a representation of a $\Gamma$-function and falls off as $Z^{-3/4}$ independent of $n$. From this, it follows that imposing a classical asymptotic behavior in the infrared does not constrain the parameters $b_n$.\footnote{
Alternatively, one could notice that the limit in eq. (\ref{IRcondition}) is formally of the same type as considered in
Appendix A of \cite{Aslanbeigi:2014zva}, and thus one can apply the same manipulations to conclude that the limit gives zero irrespective of $n$.}

\begin{figure}
	\begin{centering}
		\includegraphics[width=0.45\textwidth]{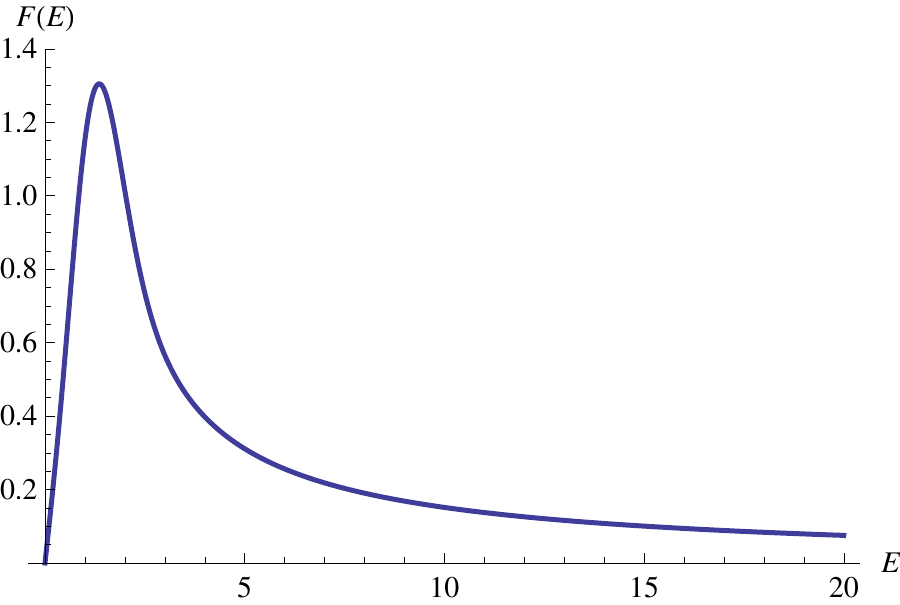} \, 
		\includegraphics[width=0.45\textwidth]{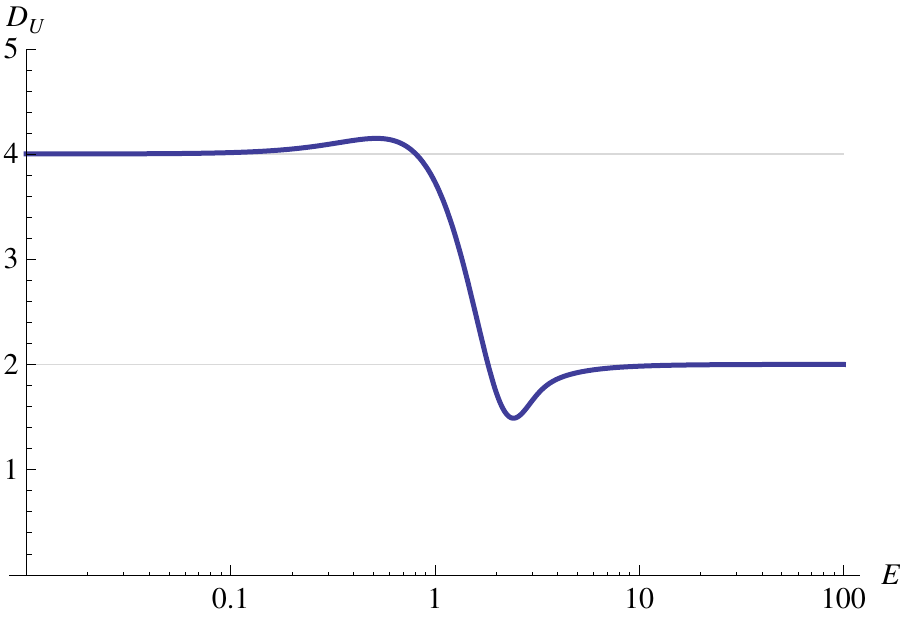}
		\caption{\label{Fig.causal} Profile function ${\cal F}(E)$ and  Unruh dimension $D_U$ arising from \eqref{csmodel} with $b_0 = 1$, $b_1 = -2$ and $b_n = 0, n \ge 2$.
		}
	\end{centering}
\end{figure}

Evaluating \eqref{spectralF} for \eqref{wfcaus2} yields the profile function for this model
\be\label{csmodel}
\cF(E) = E + \sum_{n=0}^N \, b_n \, \int_0^{E^2} dm^2 \, e^{-\alpha m^2} \, m^{2n} \, \sqrt{E^2 - m^2 } \, . 
\ee
At this stage, it is instructive to study the case $N=1$ in detail. Setting $\alpha = 1$, the two integrals can be carried out explicitly, giving rise to imaginary error functions
\be\label{int1}
\begin{split}
	I_0 \equiv & \, \int_0^{E^2}dx \, e^{-x} \, \sqrt{E^2-x} 
	 = E - \tfrac{\sqrt{\pi}}{2} \, e^{-E^2} \, {\rm Erfi}(E) \, , \\
	 	I_1 \equiv & \, \int_0^{E^2}dx \, e^{-x} \, x \, \sqrt{E^2-x} 
	 	= \tfrac{3}{2} E - \tfrac{\sqrt{\pi}}{4} e^{-E^2} \,  (3+2E^2) \, {\rm Erfi}(E) \, .
	\end{split}
	\ee
Expanding the integrals at $E = 0$ one has
\be
I_0 \simeq \tfrac{2}{3} E^3 + \ldots \, , \qquad I_1 \simeq \tfrac{4}{15} E^5 + \ldots \, . 
\ee
Thus the low-energy behavior is governed by the massless contribution, independently of the values $b_0$ and $b_1$. Looking at the asymptotics of the integrals \eqref{int1} for $E^2 \gg 1$, one has
\be
I_0 \simeq E - \frac{1}{2E} + \ldots \, , \qquad I_1 \simeq E - \frac{3}{4E} + \ldots \, . 
\ee
Hence, for generic values $b_0$, $b_1$ the asymptotic scaling for $E \ll 1$ and $E \gg 1$ is identical. In these cases there is no change in the Unruh dimension. For the special value $b_1 = - (b_0 +1)$, however, the leading term in the high-energy expansion cancels and the asymptotics of the profile function reads
\be
\cF(E) = \frac{b_0 + 3}{4E} + \ldots \, . 
\ee 
Thus, for this case the model matches the Unruh rate expected for  Causal Set Theory. Setting $b_0 = 1$ the full profile function is shown in Fig.\ \ref{Fig.causal}.
Both the profile function and the Unruh dimension undergo a transition when the energy scale meets the discretization scale controlled by setting $b_0 = 1$.

\section{Conclusions and outlook}
\label{sect.5}

In this work we investigated the Unruh effect in quantum gravity inspired models exhibiting dynamical dimensional flows.
Since both the detector approach to the Unruh effect 
and dimensional flows originate from a non-trivial momentum dependence of the two-point correlation functions there is a natural connection between the two.
Explicitly, we focused on two-point functions arising within the context of phenomenologically motivated models 
for dynamical dimensional reduction, multiscale models, Kaluza-Klein theories, spectral actions, and Causal Set Theory.
From the viewpoint of two-point functions, these models come in two distinguished classes. 
In the first case the inverse two-point function has a polynomial expansion in momentum space. 
This case is realized within dynamical dimensional reduction, multiscale models, Kaluza-Klein theories, 
and certain classes of spectral actions. 
It is also realized in theories that break Lorentz invariance, which we did not touch upon.\footnote{
There is a vast literature on this class of models. See for instance \cite{Rinaldi:2008qt,Majhi:2013koa,Gutti:2010nv,Agullo:2008qb}}
The models forming the second class possess two-point functions which are quasi-local in the sense that their inverse consists of a first order polynomial multiplying a function which is analytic in the complex plane. This setup is realized by Causal Set Theory. Our study of these models exhibits two universal features. 
First, despite incorporating quantum (gravity) corrections in the two-point function, the Unruh radiation remains thermal in all cases. Moreover, the low-energy spectrum is robust with respect to corrections of the two-point functions at high energies, i.e., the response of an Unruh detector is not modified below the characteristic scale where the dimensional flow sets in.

The two-point functions occurring in the first class of models can be reduced to a sum of (massive) second order propagators through an Ostrogradski-type decomposition. In this case we derive a master formula which expresses the response function of the Unruh detector as a function of the mass poles. As a generic feature, one finds that dynamical dimensional reduction leads to a suppression of the Unruh effect at high energies while the opening up of extra dimensions leads to an enhancement above the compactification scale. In particular, models where the spectral dimension asymptotes to $D_s \rightarrow 2$ at high energies also exhibit a universal falloff in the rate function \eqref{defresf} of the Unruh effect ${\mathcal F}(E) \propto 1/E$.
We proposed here to quantify this non-trivial asymptotic behavior of the profile function through a new
parameter, which we called the \emph{Unruh dimension} of the system.
This is defined through the scaling of the profile function, as in eq.\ (\ref{UnruhDimension}).
Differently from other proposed parameters characterizing the high energy behavior
induced by quantum gravity effects, this one is directly related to a physical quantity
that is accessible experimentally, at least in principle. Moreover, it is directly related to the spectral dimension via the relation \eqref{dimrel}.
The specific examples studied in this paper already indicate that different quantum gravity models come with a very distinguished signature in terms of their Unruh detector response function. This may serve as an interesting starting point towards identifying universal features among different approaches to quantum gravity. This requires the computation of positive-frequency Wightman functions within different quantum gravity programs. 

Obviously, it would be quite natural to apply the formalism developed in this paper to the gravitational Asymptotic Safety program \cite{Niedermaier:2006wt,Codello:2008vh,Litim:2011cp,Reuter:2012id,Reuter:2012xf,Nagy:2012ef}. In this context, the momentum dependence of two-point functions has recently been studied in \cite{Dona:2013qba,Christiansen:2015rva,Meibohm:2015twa}. 
It is clear that an investigation of the Unruh effect should be based on the renormalized propagators 
where all quantum (gravity) fluctuations have been integrated out. 
The corresponding expression for the positive-frequency Wightman function is currently not available.
Nevertheless, much progress has been made in recent years towards the construction of
renormalized two-point functions taking quantum fluctuations into account
\cite{Codello:2013fpa,Christiansen:2012rx,Christiansen:2014raa,Christiansen:2015rva,Meibohm:2015twa,Eichhorn:2016esv}.
On this basis, we expect that it is feasible to compute the fingerprints of Asymptotic Safety in the Unruh effect.
This may also be relevant for understanding the fate of black holes within Asymptotic Safety \cite{Bonanno:1998ye,Bonanno:2000ep,Bonanno:2006eu,Reuter:2006rg,Reuter:2010xb,Becker:2012js,Falls:2012nd,Becker:2012jx,Koch:2013owa,Koch:2014cqa,Saueressig:2015xua,Litim:2013gga} based on first principles.

Another natural extension of our work is the application to Hawking radiation. Here it was argued that the low-energy Hawking spectrum is actually insensitive to Planck scale effects \cite{Agullo:2009wt}. The situation is quite similar to the one encountered in the present work, where the Unruh spectrum at energy scales below the scale where the dimensional flow sets in is actually unaltered. At the same time there are indications that quantum gravity effects could stop the black hole evaporation process and leave a cold remnant. In particular, it was argued in \cite{Carlip:2011uc} that the black hole evaporation could come to an end once the spectral dimension drops to $D_s = 3$.
This would be relevant for the information problem as well \cite{Chen:2014jwq}.
Applying the techniques based on two-point correlation functions used in the present work may actually allow one
to develop these ideas based on a first-principle calculation. 
We plan to come back to this point in the near-future.

Finally, we have not analyzed the class of models displaying a minimal length.
These models are important for quantum gravity phenomenology, since this effect
is believed to appear quite generically \cite{Garay:1994en}.
It would be interesting to see if a connection to our results can be made.

\vspace{2cm}

\paragraph*{Acknowledgements.}
We thank S. Carlip for helpful discussions and J.\ Louko for comments on the first version of the manuscript. The research of F.~S., G.~D.\ and N.~A.\ 
is supported by the Netherlands Organisation for Scientific
Research (NWO) within the Foundation for Fundamental Research on Matter (FOM) grants
13PR3137 and 13VP12.

\vspace{2cm}

\appendix

\section{Uniformly accelerated frames}
\label{App.A}
Throughout this paper we make repeated use of the worldline of an accelerated observer.
For reasons of self-completeness, we will give here a brief derivation of the main formulas
referenced in the text, mainly following \cite{Krishnan:2010un}.

A uniformly accelerated observer in special relativity is an observer having constant acceleration
in the frame in which its instantaneous velocity is zero.
The coordinate transformation to the uniformly accelerated frame defines the so called \emph{Rindler frame}.

Consider a frame $K^{\prime}$ moving with velocity $v$ along the $x$ direction with respect to a reference frame $K$.
The Lorentz boost to $K^{\prime}$ is thus $t^{\prime}=\sqrt{1-v^2}(t-vx)$, $x^{\prime}=\sqrt{1-v^2}(x-vt)$,
$y^{\prime}=y$, $z^{\prime}=z$. An object with velocity $dx/dt$ in $K$ will have a relative velocity 
in $K^{\prime}$ equal to
\be
\frac{dx^{\prime}}{dt^{\prime}} = \frac{dx/dt-v}{1-v dx/dt} \, . 
\ee
The relative acceleration is then easily found to be
\be
\label{acc}
\frac{d^2x^{\prime}}{dt^{\prime\,2}} = \frac{(1-v^2)^{3/2}}{(1-v dx/dt)^3}\frac{d^2x}{dt^2} \, . 
\ee
Imposing the instantaneous velocity in $K^{\prime}$ to be zero implies $v=dx/dt$,
which substituted in eq. (\ref{acc}) gives, for constant $d^2x^{\prime}/dt^{\prime\,2}=a$, the equation
\be
\frac{d^2x}{dt^2} = a \left(  1- \left(\frac{dx}{dt} \right)^2 \right) \, . 
\ee
This is integrated to give $x(\tau)^2 - t(\tau)^2 = a^{-2}$, or
\be
x(\tau)= a^{-1} \cosh(a\tau),\;\;\; t(\tau)=a^{-1} \sinh(a\tau) \, . 
\ee

Defining $\rho=a^{-1}$ and $\eta=a\tau$, we can then go from Minkowski coordinates
$(t,x,y,z)$ to Rindler coordinates $(\eta,\rho,y,z)$, where the new line element is
\be
ds^2=\rho^2 d\eta^2 -d\rho^2 -dy^2 -dz^2 \, . 
\ee
Starting from this, the Unruh effect can also be derived using the methods of quantum field theory in curved spacetime adapted to Rindler spacetime.

\end{document}